\documentstyle[aps,prb,eqsecnum]{revtex}
\begin{document}
\draft
\title{Free Energy of an Inhomogeneous Superconductor: a Wave Function
  Approach}

\author{Ioan Kosztin,$^{1,2}$ \v{S}imon Kos,$^1$ Michael Stone$^1$ and
  Anthony J.~Leggett$^1$}

\address{$^1$Department of Physics, University of Illinois at
  Urbana-Champaign, 1110 West Green Street, Urbana, Illinois 61801\\ $^2$The
  James Franck Institute, The University of Chicago, 5640 South Ellis Ave.,
  Chicago, IL 60637}

\date{March 26, 1998}

\maketitle
\begin{abstract}
  A new method for calculating the free energy of an inhomogeneous
  superconductor is presented.  This method is based on the quasiclassical
  limit (or Andreev approximation) of the Bogoliubov--de Gennes (or wave
  function) formulation of the theory of weakly coupled superconductors.
  The method is applicable to any pure bulk superconductor described by a
  pair potential with arbitrary spatial dependence, in the presence of
  supercurrents and external magnetic field.  We find that both the local
  density of states and the free energy density of an inhomogeneous
  superconductor can be expressed in terms of the diagonal resolvent of the
  corresponding Andreev Hamiltonian, resolvent which obeys the so-called
  Gelfand--Dikii equation.  Also, the connection between the well known
  Eilenberger equation for the quasiclassical Green's function and the less
  known Gelfand--Dikii equation for the diagonal resolvent of the Andreev
  Hamiltonian is established.  These results are used to construct a general
  algorithm for calculating the (gauge invariant) gradient expansion of the
  free energy density of an inhomogeneous superconductor at arbitrary
  temperatures.
\end{abstract}

\pacs{PACS numbers: 
   74.20.-z,  
   74.20.Fg, 
   74.80.-g 
\hfill [{\tt cond-mat/9803317}]}
\section{Introduction}
\label{sec:2-intro}

The most interesting, and also most difficult, problems in the theory of
weak coupling (BCS) superconductivity \cite{bcs} are those in which the pair
potential (order parameter) has both spatial and time dependence. 
Examples of such problems are the electromagnetic response of
superconductors \cite{schrieffer}, relaxation phenomena and collective modes
in superconductors \cite{pethick79-133}, vortex motion in bulk
superconductors \cite{bardeen65-A1197,nozieres66-667,kimm-parks}, quantum
tunneling of vortices \cite{blatter91-3297}, phase slips in
quasi-one-dimensional superconducting wires
\cite{langer67-498,mccumber70-1054,giordano94-460,sharifi93-428,ling95-805},
fluctuation effects above $T_c$ \cite{azlamazov68-875}, etc.
In principle all these phenomena can be described in the framework of the
microscopic theory of BCS superconductivity in one of its formulations based
on either Green's functions \cite{agd}, or functional integrals
\cite{popov}, or the Bogoliubov--de Gennes (BdG) equations \cite{deGennes},
i.e., the wave function formulation.
Unfortunately, such an approach is impractical due to formidable technical
difficulties of solving the corresponding microscopic equations. The
existence of the two well separated energy scales in the problem, namely the
Fermi energy $E_F$ and the magnitude of the gap function $\Delta$ makes the
problem even more difficult as far as numerical calculations are concerned.
However, if we are interested only in the low energy (or long wavelength)
physics of superconductors then the significant difference between these two
energy scales allows us to employ the quasiclassical limit of the above
mentioned microscopic theories. The quasiclassical Green's function method
\cite{serene83-221} is probably the most efficient method developed so far
for solving problems involving inhomogeneous, non-equilibrium
superconductors. Nevertheless, this method has its own limitations too
(besides the fact that it is valid only on sufficiently long length and time
scales, for example, the complicated and counterintuitive boundary conditions
used in this method need to be determined from the underlying microscopic
theory, which often relies on questionable approximations).
Therefore, it is highly desirable to develop an effective theory of weak
coupling superconductivity which technically is fairly simple and at the
same time is general enough to allow for a correct description of the above
mentioned phenomena. Such an effective theory exists only close to the
critical temperature $T_c$, where the superconducting order parameter is
small, and a time dependent Ginzburg-Landau (TDGL) theory is well established
\cite{schmid66-302,abrahams66-416}.
Recently, attempts to develop a viable TDGL
theory valid at all temperatures \cite{saito89-55,saito89-85,stoof93-7979}
yielded some promising results but controversy concerning this subject
persist \cite{greiter89-903,ao90-677,aitchison95-6531,stone95-1359}. 

So far, all derivations of TDGL theories have been done by using Green's
functions and functional integrals. Although these methods are suitable for
describing inhomogeneous superconductors in the presence of impurities,
supercurrents and electromagnetic fields, they usually resort to
uncontrollable approximations during the decoupling of the higher order
Green's functions. These approximations may lead to unphysical solutions
corresponding to states which cannot be described by any wave function. In
fact, it is known that the Green's function method as is usually formulated
does not provide a complete dynamical description of the superconducting
system and, therefore, it needs to be extended by some extra criterion
(different from a variational principle) in order to eliminate the
spurious, unphysical solutions from the correct one
\cite{balian62-372,hone62-370}. 
A typical example in this respect is related to the ground state of the
superfluid He$^3$. Starting from the same BCS reduced Hamiltonian, one can
use at least two different forms (or, equivalently, decoupling schemes) for
the second order correlation function which, in general, lead to different
ground states and quasiparticle excitation spectrum: (1) Gor'kov and
Galitskii \cite{gorkov61-792} have obtained an isotropic ground state and
excitation spectrum, whereas (2) Anderson and Morel \cite{anderson61-1911},
whose approach corresponds to a BCS type of second order correlation
function, have obtained an anisotropic ground state and excitation spectrum.
Interestingly, the ground state energy corresponding to the isotropic state
is lower than the ground state energy of the anisotropic state, however, the
former does not correspond to any state wave function and therefore must be
rejected \cite{balian62-372}. Note that there exist other examples as well,
where the Green's function method can lead to an unphysical ground state
with energy smaller than the one obtained by solving the corresponding
Schr\"odinger equation \cite{balian62-372}.

In view of this fact it is natural to consider the wave function, or BdG,
formulation of weak coupling superconductivity to develop a TDGL theory.  A
first step in this respect is to derive an expression for the free energy
functional for an inhomogeneous superconductor in time independent
(stationary) situation.  Such a derivation is the subject of the present
work. In this paper we present a new method for calculating the free energy
density of an inhomogeneous superconductor by employing the quasiclassical
limit of the wave function formulation of the theory of superconductivity.
The method is applicable to any pure bulk superconductor described by a pair
potential with arbitrary spatial dependence, in the presence of
supercurrents and external magnetic field. We show that neither the
eigenvalues nor the corresponding eigenfunctions of the BdG Hamiltonian are
needed to calculate the free energy density, which can be expressed solely
in terms of the diagonal resolvent of the corresponding Andreev Hamiltonian,
resolvent which obeys the so-called Gelfand--Dikii equation \cite{dikii}.
One of the main features of our method is that it provides a rather simple
and systematic way to derive the (gauge invariant) gradient expansion of the
free energy density at arbitrary temperatures.

The BdG method has been applied previously in the literature to study the
physical properties of inhomogeneous superconductors. The first attempt in
this respect has been undertaken by the Orsay group
\cite{caroli64-307,deGennes}. They have determined by solving the BdG
equations in the quasiclassical (WKBJ or Andreev) approximation the low
energy excitations in the core of an isolated vortex. Also, de Gennes
\cite{deGennes} has shown that close to $T_c$ the BdG equations can be
solved by employing the Raylight-Schr\"odinger perturbation theory and as a
result one obtains the Ginzburg--Landau (GL) equations.
Later on, Bardeen {\em et al.\/} \cite{bardeen69-556} (BKJT) have made a
more systematic and detailed analysis of the structure of an isolated vortex
core. BKJT assumed a variational form for both the pair potential and the
vector potential and the variational parameters have been determined by
minimizing the corresponding free energy. Cleary \cite{cleary70-1039}
applied the theory of BKJT in the vicinity of superconducting transition
temperature $T_c$ and, quite surprisingly, besides the expected
Ginzburg--Landau terms in the free energy functional he obtained several
anomalous terms as well.  These findings have been received with great
interest by the superconductivity community and several authors have tried
to explain the origin of these anomalous terms \cite{jacobs70-3587}. As a
result of these research efforts it has been found that apart from the
vortex problem anomalous terms in the free energy density also appear in
other problems involving inhomogeneous superconducting systems, such as the
healing-length problem \cite{cleary70-3577,jacobs71-3247}, the N-S proximity
junction problem \cite{hu72-1}, etc. Soon after the original work of
Bar-Sagi and Kuper \cite{barsagi72-1556}, who managed to find analytically a
self-consistent solution of the BdG equations in the Andreev approximation
(i.e., the Andreev equations) by using a model pair potential
$\Delta(z)\propto\tanh(\alpha\,z)$, an intense search has been started to
discover other, practically more useful pair potentials which are
self-consistent solutions of the corresponding Andreev equations
\cite{clinton73-1962,eilenberger75-479,chen76-43}. In fact the existence of
these self-consistent pair potentials are related to the supersymmetric
property of the properly transformed Andreev Hamiltonian (see
Sec.~\ref{sec:zero_field}) where the pair potential has the role of
superpotential \cite{junker}. In can be shown that whenever the pair of
potential energies generated by the superpotential are shape invariant the
eigenstates of the corresponding supersymmetric Hamiltonians can be
determined analytically by means of simple harmonic oscillator like operator
algebra. Apparently this simple but rather important observation has not
been recognized in the literature. The problem of anomalous terms in the
gradient expansion of the free energy density has been reconsidered by Hu
\cite{hu75-3635} and Eilenberger and Jacobs \cite{eilenberger75-479} (EJ) by
using the exact self-consistent solution of certain inhomogeneous
superconducting systems. These authors demonstrated that the actual origin
of these anomalous terms are related to surface terms and terms originating
from the possible discontinuities of the pair potential or its derivatives.
EJ have also developed a beautiful theory for calculating the free energy
density of a quasi one-dimensional inhomogeneous superconductor in the clean
limit and in the absence of supercurrents and magnetic field.  Also, in a
recent work Waxman \cite{waxman94-570} starting from the Fredholm
(functional) determinant expression of the free energy of an inhomogeneous
superconductor has shown that the later can be expressed in terms of the
determinant of a finite $4\times 4$ matrix. However, no viable method for
calculating this determinant has been proposed.

The paper is organized as follows: We begin with a brief review of the BdG
method of superconductivity (Sec.~\ref{sec:2-BdG}). Next, we express the
free energy of a bulk superconductor in terms of the spectrum of the BdG
Hamiltonian and the distribution function of the quasiparticles
(Sec.~\ref{sec:2-F}). The quasiclassical (Andreev) approximation and the
expression of the free energy in this limit are presented in
Sec.~\ref{sec:2-andreev}. Next, by using the wave function formulation of
the theory of superconductivity, we describe two different methods for
calculating the free energy density and the local density of states of an
inhomogeneous superconductor. Both methods are based on expressing the free
energy density of the superconductor in terms of the diagonal resolvent of
the so-called Gelfand--Dikii equation. The first method, which is applicable
only in the absence of the magnetic field and for a real pair potential with
arbitrary spatial dependence, is presented in Sec.~\ref{sec:zero_field},
while the second method, which is more general and applicable for
superconductors in the presence of supercurrents and magnetic field, is
presented in Sec.~\ref{sec:finite_field}. Finally,
Sec.~\ref{sec:2-conclusion} is reserved for conclusions. Also, the
derivation of both {\em scalar\/} and {\em matrix\/} Gelfand--Dikii
equations, which play a key role in our calculations of the free energy, are
provided in two appendixes. 

\section{The Bogoliubov--de Gennes Equations}
\label{sec:2-BdG}

The Bogoliubov--de Gennes (BdG), or wave function, formulation of the
microscopic theory of weak coupling superconductors represents an attractive
alternative to the widely used Green's function and functional integral
methods. The BdG method is conceptually simple, requires only knowledge of
elementary quantum mechanics, yet it is general and powerful. In what
follows we apply this method to evaluate the free energy density of an
inhomogeneous conventional $s$-wave superconductor.

Mainly to establish notations, we begin with a brief review of the basic
equations of the BdG method \cite{deGennes}.
Consider a pure bulk superconductor in the presence of a static magnetic
field. The system is described by the effective mean field Hamiltonian
\cite{deGennes}
\begin{equation}
  \label{eq:2-BdG-1}
  {\cal H}_{\text{eff}} \;=\;
  \int\!d^{\,3}\bbox{r}\,\left[\psi^{\dagger}_{\sigma}(\bbox{r})\, 
  H_o(\bbox{r})\,\psi_{\sigma}(\bbox{r}) +
  \Delta(\bbox{r})\,\psi^{\dagger}_{\uparrow}(\bbox{r})\,
  \psi^{\dagger}_{\downarrow}(\bbox{r}) +
  \Delta^*(\bbox{r})\,\psi_{\downarrow}(\bbox{r})\,
  \psi_{\uparrow}(\bbox{r}) +
  \frac{\left|\Delta(\bbox{r})\right|^2}{V} 
  \right]\;,
\end{equation}
where $\Delta(\bbox{r})$ is the (mean-field) pair potential, $V$ is the
Gor'kov contact pairing interaction [i.e., $V\left(\bbox{r}-\bbox{r}'\right)
= V\,\delta\left(\bbox{r}-\bbox{r}'\right)$], the field operators
$\psi_{\sigma}(\bbox{r})$ and $\psi^{\dagger}_{\sigma}(\bbox{r})$ destroy
and create, respectively, an electron at position $\bbox{r}$ with spin
orientation $\sigma=\uparrow,\downarrow$, and obey the usual fermionic
anticommutation relations
\begin{equation}
  \label{eq:2-BdG-2}
  \left\{\psi_{\sigma}(\bbox{r}),\psi_{\sigma'}(\bbox{r}') \right\} \;=\; 0\;,
  \qquad 
  \left\{\psi^{\dagger}_{\sigma}(\bbox{r}),
  \psi^{\dagger}_{\sigma'}(\bbox{r}')\right\} \;=\; 0 \;, \qquad
  \left\{\psi_{\sigma}(\bbox{r}),
  \psi^{\dagger}_{\sigma'}(\bbox{r}')\right\} \;=\;
  \delta_{\sigma\sigma'}\,\delta\left(\bbox{r}-\bbox{r}'\right)\;,
\end{equation}
and, finally, the kinetic energy operator, measured with respect to the
Fermi energy $E_{\text{F}}$, is given by 
\begin{equation}
  \label{eq:2-BdG-3}
  H_o(\bbox{r}) \;=\;
  \frac{1}{2\,m}\left(\hat{\bbox{p}}-\frac{e}{c}\,\bbox{A}\right)^2 -
  E_{\text{F}}\;, \qquad \hat{\bbox{p}} \;=\; -i\,\hbar\nabla\;,
\end{equation}
where the vector potential $\bbox{A}(\bbox{r})$ is related to the total
magnetic field $\bbox{H}(\bbox{r})$ through the equation $\bbox{H}(\bbox{r})
= \nabla\times\bbox{A}(\bbox{r})$. 
In Eq.~(\ref{eq:2-BdG-1}), and throughout this paper, implicit summation
over repeated spin or pseudospin indices is assumed.

The effective Hamiltonian (\ref{eq:2-BdG-1}) can be diagonalized by using the
Bogoliubov transformations \cite{deGennes}
\begin{equation}
  \label{eq:2-BdG-5}
  \begin{array}{ccl}
    \psi_{\uparrow}(\bbox{r}) &=&
    \sum_i\left[u_i(\bbox{r})\,\gamma_{i\uparrow} -
      v^*_i(\bbox{r})\,\gamma^{\dagger}_{i\downarrow} \right]\;, \\ & & \\
    \psi_{\downarrow}(\bbox{r}) &=& 
    \sum_i\left[u_i(\bbox{r})\,\gamma_{i\downarrow} +
      v^*_i(\bbox{r})\,\gamma^{\dagger}_{i\uparrow} \right]\;,
  \end{array}
\end{equation}
where $i$ labels a complete set of quantum states in the relevant Hilbert
space, the $\gamma$ and $\gamma^{\dagger}$ are the Bogoliubov quasiparticle
annihilation and creation operators, respectively, and satisfy the fermionic
anticommutation rules
\begin{equation}
  \label{eq:2-BdG-6}
  \left\{\gamma_{i\alpha},\gamma_{j\beta}\right\} \;=\; 0\;, \qquad 
  \left\{\gamma_{i\alpha},\gamma^{\dagger}_{j\beta}\right\} \;=\;
  \delta_{ij}\,\delta_{\alpha\beta}\;. 
\end{equation}
The Bogoliubov amplitudes $u_i$ and $v_i$ ought to be determined by the
condition that the transformations (\ref{eq:2-BdG-5}) diagonalize ${\cal
  H}_{\text{eff}}$; they obey the so-called Bogoliubov--de Gennes (BdG)
equations \cite{deGennes} which can be written in compact form
\begin{equation}
  \label{eq:2-BdG}
  {\cal H}_{\text{BdG}}\,\Psi_i(\bbox{r}) \;\equiv\; 
  \left(
    \begin{array}{cc}
      H_o(\bbox{r}) & \Delta(\bbox{r}) \\ & \\
      \Delta^*(\bbox{r}) & -H_o^*(\bbox{r}) 
    \end{array}
  \right)\,\Psi_i(\bbox{r}) \;=\;
  E_i\,\Psi_i(\bbox{r})\;,
\end{equation}
where $\Psi_i\equiv\left(u_i,v_i\right)^T$ is a pseudo spinor in
particle-hole space. Thus, the pair potential mixes coherently the particle
and hole states and, as a result, the Bogoliubov quasiparticles have a mixed
particle and hole like character. After diagonalization (\ref{eq:2-BdG-1})
reads 
\begin{equation}
  \label{eq:2-BdG-7}
  {\cal H}_{\text{eff}} \;=\; E_g + \sum_i E_i\,\gamma^{\dagger}_{i\alpha}\,
  \gamma_{i\alpha} \;, 
\end{equation}
where the ground state energy is given by
\begin{equation}
   \label{eq:2-BdG-8}
   E_g \;=\; -2\sum_i E_i \int\!d^{\,3}\bbox{r}\,
  \left|v_i(\bbox{r})\right|^2 + \int\!d^{\,3}\bbox{r}\,
  \frac{\left|\Delta(\bbox{r})\right|^2}{V} \;.
\end{equation}
According to the expression (\ref{eq:2-BdG-7}) our system is equivalent to
an ``ideal gas'' of Bogoliubov quasiparticles with energies $E_i$, which are
the eigenvalues of the BdG equations (\ref{eq:2-BdG}). For an arbitrary pair
potential $\Delta(\bbox{r})$ the eigenvalue problem determined by
(\ref{eq:2-BdG}), subject to suitable boundary conditions, is difficult to
solve even numerically.

\section{Free Energy}
\label{sec:2-F}

By definition, the free energy is given by \cite{note1}
\begin{equation}
  \label{eq:2-F-1}
  {\cal F} \;=\; \left\langle {\cal H}_{\text{eff}} \right\rangle - T\,{\cal
    S}\; + {\cal F}_H,
\end{equation}
where ${\cal S}$ is the entropy, and 
\begin{equation}
  \label{eq:2-F-1a}
  {\cal F}_H \;=\; \int\!d^{\,3}\bbox{r}\,F_H(\bbox{r})\;, \qquad
  F_H(\bbox{r}) \;=\;
  \frac{\left|\bbox{H}(\bbox{r})-\bbox{H}_a\right|^2}{8\,\pi}\;,
\end{equation}
is the positive magnetic field exclusion energy due to the screening
supercurrents induced by the applied field $\bbox{H}_a$.
Also, we have assumed that the temperature distribution across the system is
homogeneous. The notation
$\langle\ldots\rangle\equiv\text{Tr}\left\{\hat{\rho}\ldots\right\}$
indicates the average over some statistical ensemble described by the
density matrix $\hat{\rho}$. In thermal equilibrium
\begin{equation}
  \label{eq:2-F-2}
  \hat{\rho} \;=\; \frac{\exp\left(-{\cal H}_{\text{eff}}/T\right)}{
  \text{Tr}\left\{\exp\left(-{\cal H}_{\text{eff}}/T\right)\right\}}\;.
\end{equation}
One defines the mean occupation number of level ``$i$'' corresponding to
spin orientation $\alpha$  by
\begin{equation}
  \label{eq:2-F-3}
  f_{i\alpha} \;=\; \left\langle \gamma^{\dagger}_{i\alpha}\,
  \gamma_{i\alpha} \right\rangle\;,
\end{equation}
and one assumes that there is no magnetic ordering in the system such that
both spin orientations are equally likely, i.e.,
\begin{equation}
  \label{eq:2-F-3a}
  f_i \;\equiv\; f_{i\uparrow} \;=\; f_{i\downarrow} \;.
\end{equation}
It is known that the entropy of an ideal gas of fermionic (quasi-)
particles, which is not necessarily in equilibrium, can be expressed in
terms of the mean occupation numbers $f_i$ as \cite{landau5}
\begin{equation}
  \label{eq:2-F-4}
  {\cal S} \;=\; -2\,\sum_i\left[f_i\,\ln{f_i} + \left(1-f_i\right)\,
    \ln\left(1-f_i\right) \right]\;,
\end{equation}
where the factor of two accounts for the two independent spin orientations.

Thus, inserting Eqs.~(\ref{eq:2-BdG-7},\ref{eq:2-F-3},\ref{eq:2-F-4}) into
(\ref{eq:2-F-1}), the free energy of the system, which is a functional of
the pair potential $\Delta(\bbox{r})$, the mean occupation numbers $f_i$,
and the vector potential $\bbox{A}(\bbox{r})$, can be written as
\begin{equation}
  \label{eq:2-F-5}
  {\cal F}\left[\Delta(\bbox{r}),f_i,\bbox{A}(\bbox{r})\right]\;\equiv\;
  {\cal F} \;=\; E_g + 2\,\sum_i E_i\,f_i - 2\,T\,\sum_i\left[f_i\,\ln{f_i}
    + \left(1-f_i\right)\, \ln\left(1-f_i\right) \right] + {\cal F}_H\;.
\end{equation}
In thermodynamic equilibrium one requires the free energy to be stationary
with respect to $\Delta$, $f_i$, and $\bbox{A}$. Hence, stationarity with
respect to: (i) the pair potential yields the so-called {\em gap equation}
(i.e., the self consistency condition for the pair potential)
\begin{equation}
  \label{eq:2-F-6}
  \Delta(\bbox{r}) \;=\;
  V(\bbox{r})\left\langle\psi_{\uparrow}(\bbox{r})\psi_{\downarrow}(\bbox{r})
  \right\rangle \;=\; V \sum_i u_i(\bbox{r})\,v^*_i(\bbox{r})
  \left(1-2\,f_i\right)\;,
\end{equation}
(ii) the mean occupation number of the state
$i$ (for either two spin orientations) yields the usual Fermi distribution
function
\begin{equation}
  \label{eq:2-F-7}
  f_i \;=\; \left[\exp\left(E_i/T\right)+1\right]^{-1}\;,
\end{equation}
and, (iii) the vector potential yields the Maxwell equation

\begin{equation}
  \label{eq:2-F-7a}
  \nabla\times\left(\nabla\times\bbox{A}(\bbox{r})\right) \;=\;
  \frac{4\pi}{c}\,\bbox{j}(\bbox{r})\;,
\end{equation}
where the supercurrent density is given by

\begin{equation}
  \label{eq:2-F-7b}
  \bbox{j}(\bbox{r}) \;=\; 2\frac{e}{m} \sum_i\left[f_i\, u_i^*(\bbox{r})
  \hat{\bbox{P}} u_i(\bbox{r}) + \left(1-f_i\right)\, v_i(\bbox{r})
  \hat{\bbox{P}} v_i^*(\bbox{r})\right]\;, \qquad \hat{\bbox{P}} \;\equiv\;
  \hat{\bbox{p}}-\frac{e}{c}\,\bbox{A}(\bbox{r})\;.  
\end{equation}
In the absence of the magnetic field, the BdG equations (\ref{eq:2-BdG})
together with Eqs.~(\ref{eq:2-F-6}) and (\ref{eq:2-F-7}) yield the standard
BCS result corresponding to a uniform and real pair potential
$\Delta(\bbox{r})=\Delta_o$; the eigenstates $i$ are plane wave states
$|\bbox{k}\rangle$ and
\begin{eqnarray}
  \label{eq:2-F-BCS}
  E_{k} &=& \sqrt{\xi_{k}^2+\Delta_o^2}\;, \qquad
  \xi_k=\frac{\hbar^2\,k^2}{2m} - E_{\text{F}} \nonumber \\
  u_k &=& \sqrt{\frac{1}{2}\left(1+\frac{\xi_k}{E_k}\right)}\;, \qquad
  v_k \;=\; \sqrt{\frac{1}{2}\left(1-\frac{\xi_k}{E_k}\right)}\;, \\
  \frac{1}{V\,N_o} &=& \int_0^{\omega_c}\!d\xi_k\,
  \frac{\tanh\left(E_k/2T\right)}{E_k} 
  \;=\; 2\,\pi\,T\sum_{\omega_n>0}^{\omega_c}
  \left(\omega_n^2+\Delta_o^2\right)^{-1/2} \;,\nonumber
\end{eqnarray}
where $N_o$ is the normal state density of states (for both spin
orientations) at the Fermi level, $\omega_c$ is a cut-off frequency of the
order of the Debye frequency, and $\omega_n=\pi\,T(2n+1)$ are fermionic
Matsubara frequencies.

In what follows we will be interested in calculating the free energy ${\cal
  F}$ for a spatially varying pair potential and magnetic field which do not
necessarily obey the self consistency equations (\ref{eq:2-F-6}) and
(\ref{eq:2-F-7a}-\ref{eq:2-F-7b}). For the moment, we assume that the
relation (\ref{eq:2-F-7}) is valid but, later on, we will relax this
condition as well (see Sec.~\ref{sec:2-noneq}). So, we consider a
superconductor in which the quasiparticles are in thermal equilibrium but
the pair potential and the magnetic field may have an arbitrary spatial
variation.  In this case the expression of the free energy can be further
simplified.
Inserting (\ref{eq:2-F-7}) into (\ref{eq:2-F-5}), and by taking into account
(\ref{eq:2-BdG-8}), after some straightforward algebra one obtains
\begin{equation}
  \label{eq:2-F-8}
  {\cal F} \;=\; -2\,T\sum_i\ln\left(2\,\cosh\frac{E_i}{2\,T}\right) +
  \int\!d^{\,3}\bbox{r}\, \frac{\left|\Delta(\bbox{r})\right|^2}{V} + 
  {\cal F}_H\;.
\end{equation}
Apparently, in order to calculate the free energy (\ref{eq:2-F-8}) it is
necessary to know the spectrum $\left\{E_i\right\}$ of the BdG Hamiltonian
${\cal H}_{\text{BdG}}$ for a given pair potential $\Delta(\bbox{r})$ (and
boundary condition).  Fortunately, this is not the case as several authors
have already shown \cite{waxman94-570,eilenberger75-479}, albeit in the
absence of any magnetic field and by assuming that $\Delta(\bbox{r})$
depends only on a single spatial coordinate. Indeed, by employing the
identity \cite{g&r}
\begin{equation}
  \label{eq:2-F-9}
  \cosh^2\left(\frac{x}{2}\right) \;=\; \prod_{m=-\infty}^{\infty}
  \left[1+\frac{x^2}{\pi^2\,(2\,m+1)^2}\right]\;,
\end{equation}
the free energy (\ref{eq:2-F-8}) can be recast as
\begin{equation}
  \label{eq:2-F-10}
  {\cal F} \;=\; -2\,T\,\sum_i\ln\left(2\,
    \prod_{\omega_m>0}\frac{\omega_m^2+E_i^2}{\omega_m^2}\right) +
  \int\!d^{\,3}\bbox{r}\, \frac{\left|\Delta(\bbox{r})\right|^2}{V} +
  {\cal F}_H\;,
\end{equation}
where $\omega_m$ are fermionic Matsubara frequencies. The formal divergence
of the above expression of the free energy can be eliminated by subtracting
from ${\cal F}$ (i.e., by measuring ${\cal F}$ with respect to) the free
energy ${\cal F}_N$ of the corresponding normal state. Thus, by denoting
$\delta{\cal F}\equiv{\cal F}-{\cal F}_N$, we have
\begin{eqnarray}
  \label{eq:2-F-12}
  \delta{\cal F} &=& -2\,T \sum_i\,\ln
  \prod_{\omega_m>0}\frac{\omega_m^2+E_i^2}{\omega_m^2+\epsilon_i^2} +
  \int\!d^{\,3}\bbox{r}\,
  \frac{\left|\Delta(\bbox{r})\right|^2}{V} + {\cal F}_H
  \nonumber \\ &=& -2\,T\,\sum_{\omega_m>0}
  \ln\text{Det}\left(\frac{\omega_m^2+ {\cal H}^2_{\text{BdG}}}{\omega_m^2+
      {\cal H}^2_o}\right) + \int\!d^{\,3}\bbox{r}\,
  \frac{\left|\Delta(\bbox{r})\right|^2}{V} + {\cal F}_H\;,
\end{eqnarray}
where, ${\cal H}_o$ is the BdG Hamiltonian corresponding to the normal state
of the system (i.e., $\Delta=0$), and $\left\{\epsilon_i\right\}$ denote the
spectrum of ${\cal H}_o$.

Waxman \cite{waxman94-570} has shown that the infinite Fredholm (functional)
determinant in Eq.~(\ref{eq:2-F-12}), which contains in an encapsulated form
all the information on the one-particle excitation spectrum of the
superconductor, can be expressed, at least in the case of a quasi
one-dimensional inhomogeneous superconductor and in the absence of the
magnetic field, in terms of a finite $4\times 4$ matrix $M$. However, the
actual evaluation of this matrix $M(x)$, which transports eigenfunctions of
${\cal H}_{\text{BdG}}$ from $x=0$ to $x=L$ ($L$ is the size of the system
in the relevant $x$ direction) is quite complicated and analytical results
are possible only for layered systems with a piecewise constant pair
potential.
In the work by Eilenberger and Jacobs \cite{eilenberger75-479} the Fredholm
determinant is calculated in terms of a function ${\cal E}(x)$ which obeys
an integral equation of Volterra type. This method seems to be somewhat
simpler than Waxman's and allows for analytical results (in the
quasiclassical limit) in several nontrivial cases and, furthermore, provides a
viable procedure to obtain the gradient expansion of the free energy density
about its equilibrium value.

In contrast to both above mentioned methods, which are only applicable when
$\Delta(\bbox{r})$ varies along a given direction, in the absence of any
external field, and with the Bogoliubov quasiparticles in thermal
equilibrium with the superconducting condensate, our method of calculating
the free energy of an inhomogeneous superconductor is valid for an arbitrary
$\Delta(\bbox{r})$, in the presence of an arbitrary static magnetic field,
and it can be generalized for an arbitrary distribution function $f_i$ of
the quasiparticles. Our method is based on the {\em quasiclassical
  approximation} of the BdG equations which we describe next.

\section{Quasiclassical (Andreev) Approximation}
\label{sec:2-andreev}

Superconductors are characterized by two different energy scales,
namely the Fermi energy $E_{\text{F}}$ and the amplitude of the pair
potential (gap function) $\Delta_o$ at zero temperature. The length scales
corresponding to these energies are the Fermi wavelength
$\lambda_{\text{F}}\sim k_{\text{F}}^{-1}\sim
\hbar\,v_{\text{F}}/E_{\text{F}}$ which gives the mean inter-particle
distance in the system, and the superconducting coherence length
$\xi_o\sim\hbar\,v_{\text{F}}/\Delta_o$ which determines the spatial extent
of the pair correlation. Since in conventional superconductors
$E_{\text{F}}\gg\Delta_o$ (or $\lambda_{\text{F}}\ll\xi_o$), as long as we
are interested only in the low energy (or long wavelength) properties of the
system it is legitimate to employ the quasiclassical approximation of the
theory of superconductivity. The BdG equations are valid on atomic scale and
therefore the spinor wave functions $\Psi_i(\bbox{r})$, which vary on a
length scale set by $k_{\text{F}}^{-1}$, contain more information than it is
necessary to calculate, for example, the free energy and free energy density
of an inhomogeneous superconductor. In general, this excess of information
is eliminated at the end of the calculations by integrating out the
irrelevant high energy (of rapidly oscillating) degrees of freedom. A more
practical approach is, however, to eliminate these irrelevant degrees of
freedom right at the beginning of the calculations by replacing the BdG
equations by their quasiclassical limit, i.e., the so-called {\em Andreev
  equations\/} \cite{andreev64-1228}. For this purpose, one writes the
spinor wave function $\Psi_i$ as a rapidly oscillating phase factor (which
changes on atomic length scale) times a slowly varying amplitude (which
changes on a length scale set by the coherence length), i.e.\cite{note2},
\begin{equation}
  \label{eq:2-andreev-1}
  \Psi_i(\bbox{r}) \;\approx\; \Phi_n(\bbox{r};\hat{\bbox{u}})\,
  \exp\left(i\,k_{\text{F}}\,\hat{\bbox{u}}\,\bbox{r}\right)\;.
\end{equation}
Thus, in the quasiclassical approximation, the quasiparticles are moving
along classical trajectories which are straight lines determined by the unit
vector $\hat{\bbox{u}}$ and the ``impact parameter'' $r_{\perp}$ (which gives the
distance of the quasiclassical trajectory from the origin of the coordinate
system); the position vector in (\ref{eq:2-andreev-1}) reads
\begin{equation}
  \label{eq:2-andreev-1a}
  \bbox{r} \;=\; x\,\hat{\bbox{u}} + \bbox{r}_{\perp} \;,
\end{equation}
where the impact parameter vector $\bbox{r}_{\perp}$ is normal to
$\hat{\bbox{u}}$. Nevertheless, the motion along the quasiclassical trajectories
is quantized and the corresponding eigenstates are labeled in
(\ref{eq:2-andreev-1}) by the quantum number $n$. So, in the quasiclassical
approximation the state $i$ is specified by the quantum numbers
$(n,\hat{\bbox{u}},\bbox{r}_{\perp})$ and the trace with respect to the original
states $i$ must be evaluated according to the formula
\begin{equation}
  \label{eq:2-andreev-2}
  \sum_i\ldots \;=\; \pi\,\hbar\,v_{\text{F}}\,N_o \int\!d^2r_{\perp}
  \int\!\frac{d\Omega_{\hat{\bbox{u}}}}{4\,\pi} \sum_n \ldots \;.
\end{equation}
Furthermore, we have (for brevity we omit the arguments)
\begin{equation}
  \label{eq:2-andreev-3}
  \nabla^2\,\Psi_i \;=\; \left(\nabla^2\,\Phi_n + 2\,i\,k_{\text{F}}\,
  \hat{\bbox{u}}\nabla\,\Phi_n - k_{\text{F}}^2\,\Phi_n\right)
  \exp\left(i\,k_{\text{F}}\,\hat{\bbox{u}}\,\bbox{r}\right)\;,
\end{equation}
and therefore, by using Eq.~(\ref{eq:2-BdG-3}) in zero magnetic field, one
obtains
\begin{equation}
  \label{eq:2-andreev-4}
  H_o\,\Psi_i \;=\; -\frac{\hbar^2}{2\,m}
  \left(\nabla^2\,\Phi_n + 2\,i\,k_{\text{F}}\,\hat{\bbox{u}}\nabla\,\Phi_n\right)
  \exp\left(i\,k_{\text{F}}\,\hat{\bbox{u}}\,\bbox{r}\right) \;\approx\;
  v_{\text{F}}\,\hat{\bbox{u}}\cdot\left(\hat{\bbox{p}}\Phi_n\right)
  \,\exp\left(i\,k_{\text{F}}\,\hat{\bbox{u}}\,\bbox{r}\right)\;, 
\end{equation}
where we have neglected the term involving the Laplacian of $\Phi_n$
({\em Andreev approximation}) because
\begin{equation}
  \label{eq:2-andreev-5}
  \left|\frac{\nabla^2\,\Phi_n}{k_{\text{F}}\,
      \hat{\bbox{u}}\nabla\,\Phi_n}\right| \sim
  \left(k_{\text{F}}\,\xi_o\right)^{-1} \ll 1\;.
\end{equation}
According to the notion of {\em minimal coupling}, in finite magnetic field
in Eq.~(\ref{eq:2-andreev-4}) one needs to replace $\hat{\bbox{p}}$ with
$\hat{\bbox{P}} = \hat{\bbox{p}} - (e/c)\,\bbox{A}$.

Note that condition (\ref{eq:2-andreev-5}) may not hold for a small fraction
of the total number of quasiclassical trajectories characterized by
$\hat{\bbox{u}}$ oriented almost perpendicular to $\nabla\Phi_n$.
The Andreev approximation also fails in spatial regions where the pair
potential (and/or its derivatives) has discontinuities, e.g., at interfaces,
boundaries etc. These non-analyticities in $\Delta(\bbox{r})$ reflect the
fact that in such regions the pair potential changes rapidly on atomic
scale. Within the quasiclassical approximation this kind of behavior of
$\Delta(\bbox{r})$ can be described by (nonintuitive) effective boundary
conditions which must be derived starting from the underlying microscopic
theory which is valid on atomic scale. It seems to be well established by
now that if one does not account properly for the possible discontinuities
in the pair potential (and or its derivatives) these can lead to unphysical
{\em anomalous} terms in the corresponding Ginzburg--Landau free energy
functional \cite{cleary70-1039,hu72-1,eilenberger75-479}.

Finally, inserting (\ref{eq:2-andreev-1}) into the BdG equations
(\ref{eq:2-BdG}) and by taking into account (\ref{eq:2-andreev-4}), one
arrives at the so-called Andreev equations \cite{andreev64-1228}
\begin{equation}
  \label{eq:2-andreev}
  {\cal H}_{\text{A}}\,\Phi_n(\bbox{r}) \;\equiv\; 
  \left(
    \begin{array}{cc}
      H(x) & \Delta(x;\hat{\bbox{u}},\bbox{r}_{\perp}) \\ & \\
      \Delta^*(x;\hat{\bbox{u}},\bbox{r}_{\perp}) & -H^*(x) 
    \end{array}
  \right)\,\Phi_n(x;\hat{\bbox{u}},\bbox{r}_{\perp}) \;=\;
  E_n(\hat{\bbox{u}},\bbox{r}_{\perp})\,\Phi_n(x;\hat{\bbox{u}},\bbox{r}_{\perp})\;,
\end{equation}
where
\begin{equation}
   \label{eq:2-andreev-a}
   H \;\equiv\; H(x) \;=\; v_{\text{F}}\,\hat{\bbox{u}}\cdot
   \left(\hat{\bbox{p}}-\frac{e}{c}\,\bbox{A}\right) 
   \;=\;  -i\,\hbar\,v_{\text{F}}\,\partial_x  - v_{\text{F}}\,\frac{e}{c}\,
   \hat{\bbox{u}}\cdot\bbox{A}(x) \;.
\end{equation}
Note that the Andreev equations (\ref{eq:2-andreev},\ref{eq:2-andreev-a})
are effectively one-dimensional; the independent variable is $x$ (the
position along the quasiclassical trajectory), and the other degrees of
freedom $\left(\hat{\bbox{u}},\bbox{r}_{\perp}\right)$ enter the equation only as
parameters. This is a key observation which allows us to treat inhomogeneous
superconductors characterized by a pair potential with arbitrary spatial
dependence.

In terms of the energy spectrum of the Andreev Hamiltonian ${\cal
  H}_{\text{A}}$ the free energy $\delta{\cal F}$ can be written as [cf.\
Eq.~(\ref{eq:2-F-12})] 
\begin{eqnarray}
  \label{eq:2-andreev-6}
  \delta{\cal F} &=& -2\,T\,\pi\,\hbar\,v_{\text{F}}\,N_o
  \int\!d^2r_{\perp} 
  \int\!\frac{d\Omega_{\hat{\bbox{u}}}}{4\,\pi} \sum_n\,\ln
  \prod_{\omega_m>0}\frac{\omega_m^2+
  E_n^2(\hat{\bbox{u}},\bbox{r}_{\perp})}{\omega_m^2+
  \epsilon_n^2(\hat{\bbox{u}},\bbox{r}_{\perp})} +  
  \int\!d^{\,3}\bbox{r}\, \frac{|\Delta(\bbox{r})|^2}{V}
  + {\cal F}_H\nonumber \\ &&\nonumber \\
  &=& -2\,T\,\pi\,\hbar\,v_{\text{F}}\,N_o \int\!d^2r_{\perp}
  \int\!\frac{d\Omega_{\hat{\bbox{u}}}}{4\,\pi}\sum_{\omega_m>0}
  \ln\text{Det}\left(\frac{\omega_m^2+ {\cal H}^2_{\text{A}}}{\omega_m^2+
      {\cal H}^2_o}\right) + \int\!d^{\,3}\bbox{r}\,
  \frac{|\Delta(\bbox{r})|^2}{V} + {\cal F}_H\;.
\end{eqnarray}
In the above expression of the free energy the Fredholm determinant involves
only the quantum states along an individual quasiclassical trajectory.

In what follows we derive a relatively simple formula for calculating the
logarithm of the above Fredholm determinant and, consequently, the free
energy. 
We begin with the case of an inhomogeneous superconductor in the absence of
supercurrents and magnetic field, where the Andreev equations can be
decoupled and, therefore, the calculations are fairly simple.
The more complicated case of a superconductor in the presence of the
magnetic field and supercurrents requires a completely new method for
calculating the free energy density. This method is presented in
Sec.~\ref{sec:finite_field}. 

\section{Superconductor in Zero Magnetic Field}
\label{sec:zero_field}

\subsection{Free Energy}

A key step in our derivation of the free energy of an inhomogeneous
superconductor in zero magnetic field is the observation, due originally to
Bar-Sagi and Kuper \cite{barsagi72-1556,barsagi74-73}, that the square of
the Andreev Hamiltonian (\ref{eq:2-andreev},\ref{eq:2-andreev-a}) can be
diagonalized and, therefore, the corresponding Andreev equations for the
spinor wave function $\Phi_n$ decouple into two independent Schr\"odinger
like equations. Indeed, by dropping all the arguments for brevity, and
assuming without any loss of generality a real pair potential, one can write
\begin{equation}
  \label{eq:2-decoupling-1}
  \Omega_{\text{A}} \;\equiv\; {\cal H}^2_{\text{A}} 
  \;=\;
   \left(
    \begin{array}{cc}
      H^2+\Delta^2 & \left[H,\Delta\right] \\ & \\
      - \left[H,\Delta\right] & H^2+\Delta^2
    \end{array}
  \right) 
  \;=\; 
  \left(
    \begin{array}{cc}
      -\hbar^2\,v^2_{\text{F}}\,\partial^2_x + \Delta^2 &
      -i\,\hbar\,v_{\text{F}}\,\left(\partial_x\Delta\right) \\ & \\
      i\,\hbar\,v_{\text{F}}\,\left(\partial_x\Delta\right) & 
      -\hbar^2\,v^2_{\text{F}}\,\partial^2_x + \Delta^2
    \end{array}
    \right) \;,
\end{equation}

$\Omega_{\text{A}}$ can be brought to diagonal form by employing the unitary
transformation
\begin{equation}
  \label{eq:2-decoupling-2}
  {\cal U} \;=\; \frac{1}{\sqrt{2}}
  \left(
    \begin{array}{cc}
      1 & 1 \\ & \\
      i & -i 
    \end{array}
  \right) \;, \qquad
  {\cal U}^{\dagger} \;=\; {\cal U}^{-1} \;=\; \frac{1}{\sqrt{2}}
  \left(
    \begin{array}{cc}
      1 & -i \\ & \\
      1 & i 
    \end{array}
  \right) \;,
\end{equation}
i.e.,
\begin{equation}
  \label{eq:2-decoupling-3}
  \Omega_{\text{A}}' \;=\; {\cal U}^{-1}\,\Omega_{\text{A}}\,{\cal U} \;=\; 
  \left(
    \begin{array}{cc}
      H_+ & 0 \\ & \\
      0 & H_-
    \end{array}
  \right) \;,
\end{equation}
where 
\begin{eqnarray}
  \label{eq:2-decoupling-4}
  H_{\pm} &=& H^2 + \Delta^2 \pm \hbar\,v_{\text{F}}\,\Delta' \nonumber\\
  &=&  -\hbar^2\,v^2_{\text{F}}\,\partial^2_x + \Delta^2 \pm
  \hbar\,v_{\text{F}}\, \left(\partial_x\Delta\right) \;.
\end{eqnarray}
Thus, the spectrum of ${\cal H}^2_{\text{A}}$ is given by the combined
spectra of the two independent one-dimensional Schr\"odinger like operators
$H_{\pm}$.

It is worthwhile noticing that the Hamiltonian $\Omega_{\text{A}}'$ is
supersymmetric (SUSY) with $\Delta(x)$ playing the role of superpotential
\cite{junker}. In the language of SUSY quantum mechanics, $H_+$ and $H_-$
correspond to the fermionic and bosonic sectors, respectively, and
supersymmetry means that the interchange of these two sectors of
$\Omega_{\text{A}}'$ leaves the dynamics of the system unchanged.
The most useful properties which result from the SUSY of the Hamiltonian
$\Omega_{\text{A}}'$ can be summarized as follows
\cite{witten81-513,junker,schwabl}:

\begin{enumerate}
  
\item The Hamiltonians $H_{\pm}$ can be expressed in terms of the ladder
  operators
  \begin{equation}
    \label{eq:2-susy-1}
    Q \;\equiv\; -\hbar\,v_{\text{F}}\,\partial_x + \Delta\;, \qquad
    Q^{\dagger} \;=\; \hbar\,v_{\text{F}}\,\partial_x + \Delta\;,
  \end{equation}
  as
  \begin{equation}
    \label{eq:2-susy-2}
    H_+ \;=\; Q^{\dagger}\,Q\;, \qquad H_- \;=\; Q\,Q^{\dagger}\;.
  \end{equation}

\item The Hamiltonians $H_{\pm}$ are positive-semidefinite isospectral (up
  to a zero mode) operators, i.e.,
  \begin{equation}
    \label{eq:2-susy-3}
    H_{\pm}\,\phi_{\pm,n} \;=\; E_n^2\,\phi_{\pm,n}\;,
  \end{equation}
  where the eigenfunctions $\phi_{+,n}$ and $\phi_{-,n}$ are related through
  \begin{equation}
    \label{eq:2-susy-4}
    \phi_{-,n} \;=\; \frac{1}{\left|E_n\right|}\,Q\,\phi_{+,n}\;, \qquad
    \phi_{+,n} \;=\; \frac{1}{\left|E_n\right|}\,Q^{\dagger}\,\phi_{-,n}\;,
    \qquad \left|E_n\right| > 0\;. 
  \end{equation}
  
\item The pairing of the eigenstates of $H_{\pm}$ fails when $E_n=0$. A zero
  mode (eigenstate with zero energy) exists whenever one of the wave
  functions
  \begin{equation}
    \label{eq:2-susy-5}
    \phi_{\pm,o} \;=\; {\cal N}\,\exp\left(\pm\int^x\!dy\,\Delta(y)\right)
  \end{equation}
  is normalizable. Since at most one of the above wave functions is
  normalizable it is clear that one may have only one zero mode belonging
  to the spectrum of either $H_+$ or $H_-$. 
  Indeed, assuming, e.g., that $\phi_{+,o}$ exists, i.e.,
  $H_+\,\phi_{+,o}=0$, then 
  \[
  \left\langle\phi_{+,o}\left|H_+\right|\phi_{+,o}\right\rangle
  \;=\;
  \left\langle\phi_{+,o}\left|Q^{\dagger}\,Q\right|\phi_{+,o}\right\rangle
  \;=\; 
  \left\|Q\left|\phi_{+,o}\right\rangle\right\| \;=\; 0
  \;\Longrightarrow\;
  \phi_{-,o} \;\propto\; Q\,\phi_{+,o} \;=\; 0\;,
  \]
  and similarly in the opposite case. 
  The necessary condition that one of $\phi_{\pm,o}$ to be normalizable is
  that $\Delta(x)$ has different signs at $x=\pm\infty$ along the
  corresponding quasiclassical trajectory. While for conventional $s$-wave
  superconductors this condition is difficult to be met in zero magnetic
  field \cite{note3}, in the case of, e.g., unconventional $d$-wave
  superconductors $\Delta(x)$ can have different signs at the two opposite
  sides of a quasiclassical trajectory which connects two different lobes of
  the order parameter
  \cite{buchholtz95-1079,buchholtz95-1099,fogelstrom97-281}.
  When the zero mode is absent we say that supersymmetry is spontaneously
  broken and the ground state of $\Omega_{\text{A}}'$ is degenerate (for a
  given $\hat{\bbox{u}}$ and $\bbox{r}_{\perp}$). When the zero mode exists one
  has a good SUSY and the zero mode is the ground state of
  $\Omega_{\text{A}}'$.
  
\item Probably the most useful feature of SUSY quantum mechanics is that it
  allows us to calculate analytically both the spectrum and the
  eigenfunctions of the partner Hamiltonians $H_{\pm}$ by means of simple
  algebraic manipulations, provided that the partner potentials
  $U_{\pm}\left(x;a_o\right) \equiv \Delta^2\left(x;a_o\right)\,\pm\,\hbar\,
  v_{\text{F}}\,\Delta'\left(x;a_o\right)$ are {\em shape invariant}
  \cite{gendenshtein83-356}, i.e., when they obey the condition
  \cite{junker}
  \begin{equation}
    \label{eq:2-susy-6}
    U_+\left(x;a_o\right) \;=\; U_-\left(x;a_1\right) + R\left(a_1\right)\;,
  \end{equation}
  where $a_1$ is a new set of parameters uniquely determined from the old
  ones $a_o$ via the mapping $a_1=F\left(a_o\right)$, and the residual term
  $R\left(a_1\right)$ is independent of $x$. A few examples of
  superpotentials which yield shape invariant potentials $U_{\pm}$ are: (i)
  $\Delta\left(x;a_o\right) \propto a_o\,\tanh(\eta\,x)$, (ii)
  $\Delta\left(x;a_o\right) \propto 1+a_o\,\exp(-\eta\,x)$, (iii)
  $\Delta\left(x;a_o\right) \propto a_o/\left[1+\exp(-\eta\,x)\right]$, and
  (iv) $\Delta\left(x;a_o\right) \propto a_o(1+\eta\,x)$. For all these
  model pair potentials the eigenstates of the Andreev Hamiltonian can be
  determined analytically by using the machinery of SUSY quantum mechanics
  \cite{junker}. Once the eigenstates of ${\cal H}_{\text{A}}$ have been
  determined it is possible to evaluate numerically the value of the
  parameter $\eta$ by imposing the self-consistency condition
  (\ref{eq:2-F-6}). Successful calculations along this line have been
  reported by Bar-Sagi and Kuper \cite{barsagi72-1556,barsagi74-73} for the
  pair potential (i), by Clinton \cite{clinton73-1962} for case (ii), and by
  Eilenberger and Jacobs \cite{eilenberger75-479} for cases (iii)-(iv).
  Of course, in principle, it is possible to obtain analytical results for
  all known nontrivial superpotentials (i.e., $\Delta(x)$ in our case) which
  lead to shape invariant (or factorizable, in the language of Infeld and
  Hull \cite{infeld51-21}) potentials $U_{\pm}(x)$ with the possibility of
  even satisfying the self-consistency (gap) equation (\ref{eq:2-F-6}).
  Unfortunately none of these ``super'' pair potentials correspond to real
  physical situations and, therefore, we will not pursue here this issue in
  any further details. Nevertheless, it is fair to recognize the potential
  usefulness of the application of SUSY quantum mechanics in the study of
  inhomogeneous superconductors within the framework of the Andreev
  approximation, a fact which to our knowledge has not been fully realized
  so far in the literature.
\end{enumerate}

Before proceeding any further it is useful to introduce new length ${\cal
  L}$ and energy ${\cal E}$ units via the definitions
\begin{equation}
  \label{eq:2-units}
  {\cal L} \;\equiv\; \frac{\hbar\,v_{\text{F}}}{\Delta_o}\,\sim\,\xi_o\;,
  \qquad \text{and} \qquad 
  {\cal E} \;\equiv\; \Delta_o\;,
\end{equation}
where $\Delta_o$ is a suitably chosen constant pair potential, e.g., the
equilibrium BCS gap parameter at the considered temperature $T$.
In these new units
\begin{eqnarray}
   \label{eq:2-decoupling-4a}
   H_{\pm} &=& H^2 + \Delta^2 \pm \Delta' \nonumber\\
   &=&  -\partial^2_x + \Delta^2 \pm \partial_x\Delta\;.
\end{eqnarray}
It is also convenient to measure the free energy in units of
$N_o\,\Delta_o^2$. 
We shall use these units throughout this section.

In what follows, the fact that $H_{\pm}$ are supersymmetric will play no
special role. The important thing is that $H_{\pm}$ are independent and
Schr\"odinger like.

Now we introduce the diagonal resolvents $R_{\pm}$ of the operators
$H_{\pm}$ which will play the central role in our method for evaluating the
free energy of an inhomogeneous superconductor in zero field. By definition
\begin{equation}
  \label{eq:2-resolvent-1}
  R_{\pm}(x;\lambda) \;\equiv\;
  R_{\pm}\left(x;\lambda;\hat{\bbox{u}},\bbox{r}_{\perp}\right) \;=\; -
  \left\langle x\left|\left(\lambda-H_{\pm}\right)^{-1}
    \right|x\right\rangle \;.
\end{equation}
Hence
\begin{equation}
   \label{eq:2-resolvent-2}
   \int_{-\infty}^{\infty}\!dx\,R(x;\lambda) \;=\;
   -\sum_n\frac{1}{\lambda-E_n^2} \;,
\end{equation}
and a similar relation holds for $R_o$ corresponding to the reference
state described by the Hamiltonian ${\cal H}_o$. In
Eq.~(\ref{eq:2-resolvent-2}) we have used the shorthand notation
\begin{equation}
  \label{eq:2-resolvent-3}
  R \;\equiv\; R_+ + R_-\;.
\end{equation}
Next, one integrates both sides of Eq.~(\ref{eq:2-resolvent-2}) with respect
to the spectral variable $\lambda$
\begin{equation}
  \label{eq:2-resolvent-4}
  \int^{\lambda}\!d\lambda \int_{-\infty}^{\infty}\!dx\,R(x;\lambda)
  \;=\; -\sum_n\,\ln\left|\lambda-E_n^2\right| + \text{const}\;.
\end{equation}
The integration constant on the RHS of (\ref{eq:2-resolvent-4}) can be
eliminated by subtracting from this equation the one corresponding to the
reference state. By introducing the notation 
\begin{equation}
  \label{eq:2-resolvent-5}
  \delta{R} \;\equiv\; R - R_o\;,
\end{equation}
we get
\begin{equation}
  \label{eq:2-resolvent-6}
  \int_{-\infty}^{\lambda}\!d\lambda
  \int_{-\infty}^{\infty}\!dx\,\delta{R}(x;\lambda) \;=\;
  -\sum_n\,\ln\left|\frac{\lambda-E_n^2}{\lambda-\epsilon_n^2}\right|\;.
\end{equation}
Finally, by setting $\lambda=-\omega^2_m$ in this last equation, the
logarithm of the Fredholm determinant in (\ref{eq:2-andreev-6}) can be
written in terms of the diagonal resolvent $\delta{R}$ as
\begin{equation}
  \label{eq:2-resolvent-7}
  \ln\text{Det} 
  \left(
    \frac{\omega_m^2+{\cal H}^2_{\text{A}}}{\omega_m^2+{\cal H}^2_o}
  \right)
  \;=\; 
  \sum_n\ln\left(\frac{\omega^2_m+E_n^2}{\omega^2_m+\epsilon_n^2}\right) 
  \;=\;
  -\int_{-\infty}^{-\omega^2_m}\!d\lambda\int_{-\infty}^{\infty}\!dx\,
  \delta{R}(x;\lambda)\;.
\end{equation}
Thus, the free energy (\ref{eq:2-andreev-6}) becomes
\begin{equation}
  \label{eq:2-resolvent-8}
  \delta{\cal F} \;=\; 2\,T\,\pi\, \int\!d^2r_{\perp}
  \int\!\frac{d\Omega_{\hat{\bbox{u}}}}{4\,\pi}\sum_{\omega_m>0}
  \int_{-\infty}^{-\omega^2_m}\!d\lambda\int_{-\infty}^{\infty}\!dx\,
  \delta{R}\left(x;\lambda;\hat{\bbox{u}},\bbox{r}_{\perp}\right)
  + \int\!d^{\,3}\bbox{r}\,
  \frac{\left[\Delta(\bbox{r})\right]^2}{V}\;,
\end{equation}
where, for clarity, we have listed all the arguments of the diagonal
resolvent. 

Now the free energy density 

\begin{equation}
  \delta{F} \;=\; \frac{d(\delta{\cal F})}{d^{\,3}\bbox{r}} \;=\;
  \frac{d(\delta{\cal F})}{dx\,d^2r_{\perp}}\;,
\end{equation}
as a functional of the inhomogeneous pair potential, can be readily
extracted from Eq.~(\ref{eq:2-resolvent-8})
\begin{equation}
 \label{eq:2-F_dens-1} 
 \delta{F} \;\equiv\; \delta{F}[\Delta(\bbox{r})] \;=\; 
 \left\langle 2\,\pi\,T\sum_{\omega_m>0}
   \int_{-\infty}^{-\omega^2_m}\!d\lambda\,\delta{R} \right\rangle +
 \frac{\left[\Delta(\bbox{r})\right]^2}{V}\;,
\end{equation}
where $\langle\ldots\rangle = \int\!d\Omega_{\hat{\bbox{u}}}/4\,\pi\ldots$
means averaging over the directions of the quasiclassical trajectories.
Note that the only difference between the cases, when the pair potential
depends only on one coordinate and when it has an arbitrary $\bbox{r}$
dependence, is that in the former case the diagonal resolvent does not depend
on the impact parameter $\bbox{r}_{\perp}$ whereas in the latter case it does.
The above expression of the free energy density does not contain explicitly
either the eigenvalues or the eigenfunctions of the Andreev Hamiltonian
${\cal H}_{\text{A}}$. All the information about the superconductor is
encapsulated in the diagonal resolvent $\delta{R}$ which, however,
needs to be determined first in order to make (\ref{eq:2-F_dens-1}) useful.

Since $R_{\pm}$ are the diagonal resolvents of the one-dimensional
Schr\"odinger operators $H_{\pm}=-\partial^2_x + U_{\pm}$, with
$U_{\pm}=\Delta^2\,\pm\,\Delta'$, they obey the so-called {\em
  Gelfand--Dikii} equation \cite{gelfand-dikii,dikii}
\begin{equation}
  \label{eq:2-dikii}
  -2\,R_{\pm}\,R_{\pm}'' + R_{\pm}^{\prime\,2} +
  4\,R_{\pm}^2\,\left(U_{\pm}-\lambda\right) \;=\; 1\;.
\end{equation}
For completeness a simple derivation of this equation is provided in
Appendix~\ref{sec:ap1} (see also Ref.~\onlinecite{feinberg95-625}).
Equations (\ref{eq:2-F_dens-1}) and (\ref{eq:2-dikii}) tell us that the free
energy density of an inhomogeneous superconductor can be expressed solely in
terms of the solution of a nonlinear second order ordinary differential
equation. Unfortunately the Gelfand-Dikii equation cannot be solved
analytically for an arbitrary pair potential.
However, both the diagonal resolvent and the free energy density can be
calculated numerically once some appropriate boundary conditions have been
specified.
In this respect our method of calculating $\delta{F}$ is similar to the ones
considered by Waxman \cite{waxman94-570} and Eilenberger and Jacobs
\cite{eilenberger75-479}. However, while their methods are applicable {\em
  only} to superconductors described by a pair potential which depends on a
single spatial coordinate and in the absence of supercurrents and magnetic
field, our method is valid for pair potentials with {\em arbitrary\/}
spatial dependence.
Another important feature of our approach is that it provides a simple and
systematic way for obtaining the gradient expansion of $\delta{F}$ for an
inhomogeneous superconductor with $\Delta(\bbox{r})$ varying slowly on a
length scale $\ell\gg\xi_o$.
%
\subsection{Gradient Expansion}
\label{sec:2-grad}

\noindent For the normal state the pair potential $U_{\pm}=0$, and
(\ref{eq:2-dikii}) yields
\begin{equation}
  \label{eq:2-grad-1}
  R_{o,\pm} \;=\; \frac{1}{2\,\sqrt{-\lambda}}\;.
\end{equation}
For an arbitrary pair potential the general solution of the Gelfand--Dikii
equation (\ref{eq:2-dikii}) can be sought as an asymptotic series expansion
\begin{equation}
  \label{eq:2-grad-2}
  R_{\pm}(x;\lambda) \;=\; \frac{1}{2}\sum_{k=0}^{\infty}
  R_k^{\pm}(x)\,\left(\Delta^2-\lambda\right)^{-k-\frac{1}{2}}\;, 
\end{equation}
where $\Delta\equiv\Delta(x)$ is the pair potential of the inhomogeneous
superconductor.
For the uniqueness (up to a sign) of this expansion see, e.g.,
Ref.~\onlinecite{dikii}.
Equation (\ref{eq:2-grad-2}) is the main ingredient in our derivation of the
gradient expansion of $\delta{F}$. Our strategy is to express first
$\delta{F}$ in terms of $R_k^{\pm}(x)$, $k=0,1,\ldots$, and then to evaluate
these expansion coefficients. The latter task can be accomplished in a
systematic way by inserting (\ref{eq:2-grad-2}) into the Gelfand--Dikii
equation (\ref{eq:2-dikii}) and equating the coefficients of the different
integer powers of $\zeta\equiv\Delta^2-\lambda$ in the resulting equation.
Although this method can be used to derive a cumbersome analytical
expression for the recursion relation obeyed by the coefficients
$R_k^{\pm}(x)$, in practice it is more convenient to carry out the
calculations by employing a computer software which is suitable for
sophisticated symbolical calculations, such as {\em
  Mathematica\/}\cite{mathematica}.

It is easy to see that the first coefficient $R_0^{\pm}=1$.  Clearly, for
the normal state $R_{o,0}^{\pm}=1$ and the rest of the coefficients vanish
identically [cf.~Eqs.~(\ref{eq:2-grad-1}) and (\ref{eq:2-grad-2})], i.e.,
$R_{o,k}^{\pm}=0$, $k=1,2,\ldots$.
Thus, if one defines $\delta{R}_k \equiv R_k^+ + R_k^-$, $k=1,2,\ldots$, one
can write
\begin{equation}
  \label{eq:2-grad-3}
  \delta{R}(x;\lambda) \;=\; \left( \frac{1}{\sqrt{\Delta^2-\lambda}} -
  \frac{1}{\sqrt{-\lambda}} \right) +
  \frac{1}{2}\sum_{k=1}^{\infty}
  \delta{R}_k(x)\,\left(\Delta^2-\lambda\right)^{-k-\frac{1}{2}}\;, 
\end{equation}
and
\begin{eqnarray}
  \label{eq:2-grad-4}
  \int_{-\infty}^{-\omega_m^2}\!d\lambda\,\delta{R}(x;\lambda) &\;=\;&
  \int_{-\infty}^{-\omega_m^2}\!d\lambda\,\left(
    \frac{1}{\sqrt{\Delta^2-\lambda}} - \frac{1}{\sqrt{-\lambda}} \right) +
  \frac{1}{2}\sum_{k=1}^{\infty} \delta{R}_k(x)
  \int_{-\infty}^{-\omega_m^2}\!\frac{d\lambda}{
    \left(\Delta^2-\lambda\right)^{k+\frac{1}{2}}}   \nonumber \\ [1ex]
  &\;=\;& 2\,\left(\left|\omega_m\right| - \sqrt{\omega_m^2+\Delta^2}\right) +
  \frac{1}{2}\sum_{k=1}^{\infty}
  \frac{\delta{R}_k(x)}{\left(k-\frac{1}{2}\right)
    \left(\omega_m^2+\Delta^2\right)^{k-\frac{1}{2}}}\;.
\end{eqnarray}
Inserting (\ref{eq:2-grad-4}) into Eq.~(\ref{eq:2-F_dens-1}) yields
\begin{eqnarray}
  \label{eq:2-grad-5}
  \delta{F} &=& 4\,\pi\,T\,\sum_{\omega_m>0} \left(\left|\omega_m\right| -
    \sqrt{\omega_m^2+\Delta^2}\right) +
  \frac{\Delta^2}{V} \nonumber \\[1ex]
  &\;+\;& \left\langle\sum_{k=1}^{\infty} \left[2\,\pi\,T\sum_{\omega_m>0}
      \left(\omega_m^2+\Delta^2\right)^{-k+\frac{1}{2}}\right]\,
    \frac{\delta{R}_k(x)}{2\,k-1} \right\rangle \;.
\end{eqnarray}
The first two terms on the RHS of (\ref{eq:2-grad-5}) give the well
known bulk term contribution to the free energy density of the
superconducting state with respect to the normal state, while the third term
gives the actual gradient expansion in term of asymptotic power series of
the derivatives of the real pair potential $\Delta(x)$.

Following the above mentioned strategy for calculating the expansion
coefficients $R_{\pm}$, we wrote a {\em Mathematica\/} code which evaluates
analytically, in a systematic fashion, these coefficients. Here we apply our
results to calculate the gradient expansion of $\delta{F}$ up to the fourth
order terms, i.e.,

\begin{equation}
  \label{eq:grad-1}
  \delta{F} \;\approx\; \delta{F}_o + \delta{F}_2 + \delta{F}_4 \;,
\end{equation}
where $\delta{F}_o$ is given by the first two terms on the RHS of
Eq.~(\ref{eq:2-grad-5}). Since $\delta{R}_1=0$ there is no first order
correction to $\delta{F}$. In fact one can easily show, based on symmetry
arguments, that all odd order contributions to the gradient expansion
vanishes identically. This does not mean, of course, that all odd order
expansion coefficients $\delta{R}_{2k+1}$ are equal to zero.

To calculate $\delta{F}_2$ one needs the following coefficients
\begin{mathletters}
  \label{eq:r23}
\begin{eqnarray}
  \delta{R}_2 &\;=\;& \frac{1}{4} \left(\Delta^{\,2} -
    2\,\Delta\,\Delta''\right) \label{eq:r2} \\ [1ex]
  \delta{R}_3 &\;=\;& \frac{1}{16} \left(20\,\Delta^2\,\Delta'^{\,2} -
    \Delta''^{\,2} + 2\,\Delta'\,\Delta^{(3)} -
    2\,\Delta\,\Delta^{(4)}\right) \;.
\end{eqnarray}
\end{mathletters}
Note that while $\delta{R}_2$ contains only terms of second order in the
small parameter $\xi_o/\ell$, the coefficient $\delta{R}_3$ contains both
second and fourth order terms as well. None of the higher order coefficients
$\delta{R}_k$ contain other second order terms in $\xi_o/\ell$.  One of the
main features of our method is that it can automatically collect all the
terms of a given order in the various relevant expansion coefficients
$\delta{R}_k$.
Inserting all the second order terms from Eqs.~(\ref{eq:r23}) into
(\ref{eq:2-grad-5}) one obtains

\begin{equation}
  \label{eq:grad-f2a}
  \delta{F}_2 \;=\; \frac{1}{12}\,\left\langle 2\,\pi\,T\sum_{\omega_m>0}
  \left[\frac{\Delta'^{\,2} -
  2\,\Delta\,\Delta''}{\left(\omega_m^2+\Delta^2\right)^{3/2}} +
  5\,\frac{\Delta^2\,\Delta'^{\,2}}{\left(\omega_m^2+\Delta^2\right)^{5/2}}
  \right]\right\rangle \;.
\end{equation}
One can easily see that the second term (proportional to $\Delta''$) on the
RHS of Eq.~(\ref{eq:grad-f2a}) can also be expressed in terms of
$\Delta'^{\,2}$. Indeed, we have

\begin{equation}
  \label{eq:grad-trick}
  \frac{\Delta\,\Delta''}{\left(\omega_m^2+\Delta^2\right)^{3/2}} \;=\;
  \frac{\Delta}{\left(\omega_m^2+\Delta^2\right)^{3/2}}\,\frac{d\Delta'}{dx}
  \;=\; \frac{\,d}{dx}\left[
    \frac{\Delta\,\Delta'}{\left(\omega_m^2+\Delta^2\right)^{3/2}}\right] -
  \frac{\Delta'^{\,2}}{\left(\omega_m^2+\Delta^2\right)^{3/2}} +
  \frac{3\,\Delta^2\,\Delta'^{\,2}}{\left(\omega_m^2+\Delta^2\right)^{5/2}}
  \;.  
\end{equation}
The total derivative on the RHS yields a surface term upon integration with
respect to $x$ which, for a bulk superconductor with natural boundary
conditions, vanishes. In more complex superconductivity problems such
surface terms may lead to anomalous terms in the free energy functional
\cite{eilenberger75-479}. Nevertheless, it is important to notice that it is
always possible to express the gradient expansion of the free energy density
in terms of even powers of the pair potential and its derivatives. Another
virtue of the computer implementation of our method is that it can
automatically perform these partial integrations and return the final result
for $\delta{F}_k$ in the desired form. 

Thus, Eq.~(\ref{eq:grad-f2a}) can be rewritten as

\begin{equation}
  \label{eq:grad-f2b}
  \delta{F}_2 \;=\; \frac{1}{4}\,\left\langle 2\,\pi\,T\sum_{\omega_m>0}
  \left[\frac{\Delta'^{\,2}}{\left(\omega_m^2+\Delta^2\right)^{3/2}} -
  \frac{\Delta^2\,\Delta'^{\,2}}{\left(\omega_m^2+\Delta^2\right)^{5/2}}
  \right]\right\rangle \;.
\end{equation}

Next, we perform the average over $\hat{\bbox{u}}$, i.e., the directions of the
quasiclassical trajectories; the relevant expression is
\begin{eqnarray}
  \label{eq:2-grad-12}
  \left\langle f(\Delta)\,\Delta'^{\,2}\right\rangle &=& \left\langle
    f(\Delta)\,\left(
      \hat{\bbox{u}}\cdot\nabla\,\Delta\right)^2\right\rangle \;=\;
  \left\langle n_i\,n_j\right\rangle\,
  f(\Delta)\,\left(\partial_i\Delta\right)
  \left(\partial_j\Delta\right) \nonumber \\
  &=& \frac{1}{3}\,f(\Delta)\,\Delta_{ij}\,
  \left(\partial_i\Delta\right)\left(\partial_j\Delta\right) \;=\;
  \frac{1}{3}\,f(\Delta)\,(\nabla\,\Delta)^2 \;,
\end{eqnarray}
where $f(\Delta)$ is an arbitrary function of the pair potential.
This last result clearly depends on dimensionality; in $d$-dimensions
$\left\langle n_i\,n_j\right\rangle = \frac{1}{d}\,\delta_{ij}$.
Inserting the above results into (\ref{eq:grad-f2b}) one obtains

\begin{equation}
  \label{eq:grad-f2}
  \delta{F}_2 \;=\; \frac{1}{12}\,\pi\,T\,N_o\,\left(\hbar\,v_F\right)^2 \,
  \sum_{\omega_m>0}
  \frac{\omega_m^2}{\left(\omega_m^2+\Delta^2\right)^{5/2}}\,
  \left(\nabla\,\Delta\right)^2  \;,
\end{equation}
where we have used the original units.  This expression coincides with the
well known Werthamer result (Eq.~(129) in Ref.~\onlinecite{werthamer-parks})
for a clean superconductor in the absence of supercurrents and magnetic
field, obtained by means of many-body Green's functions.

The complexity of calculating the successive terms in the gradient expansion
of $\delta{F}$ increases exponentially with the order of the term.
Nevertheless, by using the computer implementation of our method we were
able to compute in matter of minutes the fourth order term $\delta{F}_4$.
For this purpose one needs to evaluate the expansion coefficients
$\delta{R}_k$, $k=3,\ldots,6$ and then filter out all the fourth order terms
in the small parameter $\xi_o/\ell$. After collecting all these terms, we
obtain the following expression for the fourth order term in the gradient
expansion of $\delta{F}$

\begin{equation}
  \label{eq:grad-f4a}
  \delta{F}_4 \;=\; \frac{1}{16}\left\langle 2\,\pi\,T\sum_{\omega_m>0}
    \left[\frac{7}{4}\,\left( 
        \frac{5\,\Delta^2\,\omega_m^2}{\left(\omega_m^2+\Delta^2\right)^{11/2}} 
      - \frac{\omega_m^2}{\left(\omega_m^2+\Delta^2\right)^{9/2}} 
    \right) \Delta'^{\,4}  - 
    \frac{\omega_m^2}{\left(\omega_m^2+\Delta^2\right)^{7/2}} \, \Delta''^{\,2}
  \right]\right\rangle \;.
\end{equation}
This result has been obtained after dropping irrelevant total derivatives in
order to express the final result only in terms of even powers of the first
and second derivatives of $\Delta$, the only ones which contribute to the
fourth order term in the gradient expansion. To obtain the final expression
for $\delta{F}_4$ all we need to do is to average over the directions of the
quasiclassical trajectories and to restore the original units.

Apparently, in the derivation of the results presented so far, the
particular form of the Fermi--Dirac distribution function of the
quasiparticles together with formula (\ref{eq:2-F-9}) were crucial. In what
follows we show that this is not the case and that our method of evaluating
the free energy of an inhomogeneous superconductor can be formulated in a
more general form which is also applicable for a non-equilibrium
distribution function $f_i$ of the quasiparticles. The basic idea is to
express the free energy density in terms of the {\em local density of
  states} corresponding to the Andreev Hamiltonian (i.e., along an
individual quasiclassical trajectory).

\subsection{Local Density of States}
\label{sec:2-dos}

In this section we present an alternative derivation of the expressions
(\ref{eq:2-grad-5}) of the free energy density for the case of thermal
equilibrium without invoking formula (\ref{eq:2-F-9}) but rather rewriting
the summation (\ref{eq:2-andreev-2}) over the complete set of states $i$ as
\begin{equation}
  \label{eq:2-dos-1}
  \sum_i\ldots \;=\; \pi\,\hbar\,v_{\text{F}}\,N_o \int\!d^2r_{\perp}
  \int\!\frac{d\Omega_{\hat{\bbox{u}}}}{4\,\pi} \sum_n \ldots \;=\;
  \pi\,\hbar\,v_{\text{F}}\,N_o \int\!d^2r_{\perp}
  \left\langle\int_0^{\infty}\!\rho(E)\,dE\,\ldots\right\rangle
  \;,
\end{equation}
where the density of states (DOS) along the quasiclassical trajectory
determined by $\left(\hat{\bbox{u}},\bbox{r}_{\perp}\right)$ is given by
\begin{equation}
  \label{eq:2-dos-2}
  \rho(E) \;\equiv\; \rho\left(E;\hat{\bbox{u}},\bbox{r}_{\perp}\right) \;=\;
  \sum_n \delta\left(E-E_n\left(\hat{\bbox{u}},\bbox{r}_{\perp}\right)\right)\;.
\end{equation}
Here $\left\{E_n\right\}$ represent the energy spectrum of the Andreev
Hamiltonian.
Next, let us define the DOS corresponding to the SUSY Hamiltonians $H_{\pm}$
\begin{equation}
  \label{eq:2-dos-3}
  \tilde{\rho}(E) \;=\; \rho_+(E)+\rho_-(E) \;\equiv\; \sum_n
  \delta\left(E^2-E_n^2\right) \;=\; \frac{1}{2\,E}\,\sum_n
  \delta\left(E-E_n\right) \;=\;
  \frac{\rho(E)}{2\,E}\;.
\end{equation}
Thus, from Eqs.~(\ref{eq:2-dos-3}) and (\ref{eq:2-resolvent-2}), by
employing the formula 
\[
\frac{1}{x\pm i\,0^+}\;\equiv\; \lim_{\varepsilon\rightarrow
  0^+}\frac{1}{x\pm i\,\varepsilon}=\mp 
i\,\pi\,\delta(x) + {\cal P}\,\frac{1}{x}\;,
\]
the DOS $\rho(E)$ can be expressed in terms of the diagonal resolvent
$R=R_++R_-$ as
\begin{eqnarray}
  \label{eq:2-dos-4}
  \rho(E) &=& 2\,E\,\tilde{\rho}(E) \;=\; -\frac{2\,E}{\pi}\,\text{Im}\,\sum_n
  \frac{1}{E^2-E_n^2+i\,0^+} \nonumber\\
  & & \\
  &=& \frac{2\,E}{\pi}\,\lim_{\varepsilon\rightarrow 0^+} \text{Im}\,
  \int_{-\infty}^{\infty}\!dx\, R\left(x;E^2+i\varepsilon\right)\;.
  \nonumber
\end{eqnarray}
Now combining equations (\ref{eq:2-F-8}), (\ref{eq:2-dos-1}) and
(\ref{eq:2-dos-4}) lead us to the following expression for the free energy
density 
\begin{eqnarray}
  \label{eq:2-dos-5}
  F &=& -2\,\pi\,T \int_0^{\infty}\!dE\,\rho(E;\bbox{r})\,
  \ln\left(2\,\cosh\frac{E}{2\,T}\right) + \frac{\Delta^2}{V} \nonumber \\ 
  & & \\
  &=& -4\,T\, \lim_{\varepsilon\rightarrow 0^+}
  \text{Im}\,\int_0^{\infty}\!E\,dE\, \ln\left(2\,\cosh\frac{E}{2\,T}\right)
  \left\langle R\left(x;E^2+i\,\varepsilon\right) \right\rangle
  + \frac{\Delta^2}{V} \;,\nonumber
\end{eqnarray}
where, by definition, the local DOS is given by
\begin{mathletters}
\begin{equation}
  \label{eq:2-dos-6}
  \rho(E;\bbox{r}) \;\equiv\; \frac{2\,E}{\pi}\;\text{Im}\,
  \left\langle R\left(x;E^2+i\,0^+\right) \right\rangle
  \;,
\end{equation}
or in conventional units [cf.\ Eq.~(\ref{eq:2-dos-1})]
\begin{equation}
  \label{eq:2-dos-6a}
  \rho(E;\bbox{r}) \;=\; 2\,\hbar\,v_{\text{F}}\,N_o\,E\;\text{Im}\,
  \left\langle R\left(x;E^2+i\,0^+\right) \right\rangle
  \;.
\end{equation}
\end{mathletters}

The next step is to subtract from (\ref{eq:2-dos-5}) the free energy density
corresponding to the reference normal state and to replace $\delta{R}$ in
the resulting expression by its asymptotic series expansion
(\ref{eq:2-grad-3}), i.e.,
\begin{equation}
  \label{eq:2-dos-7}
  \delta{R}\left(x;E^2+i\,\varepsilon\right) \;=\; \left(
  \frac{1}{\sqrt{\Delta^2-E^2-i\,\varepsilon}} -
  \frac{1}{\sqrt{-E^2-i\,\varepsilon}} \right) +
  \frac{1}{2}\sum_{k=1}^{\infty} 
  \delta{R}_k(x)\,\left(\Delta^2-E^2-i\,\varepsilon\right)^{-k-\frac{1}{2}}\;.
\end{equation}
Thus, the free energy density $\delta{F}=F-F_N$ becomes
\begin{equation}
  \label{eq:2-dos-8}
  \delta{F} \;=\; \delta{F}_o  
  -2\,T\,\sum_{k=1}^{\infty}\lim_{\varepsilon\rightarrow 0^+}
  \text{Im}\,\int_0^{\infty}\!E\,dE\,
  \frac{\ln\left(2\,\cosh\frac{E}{2\,T}\right)}{
  \left(\Delta^2-E^2-i\,\varepsilon\right)^{k+\frac{1}{2}}}  
  \left\langle\delta{R}_k(x)\right\rangle \;,
\end{equation}
where, the zeroth order term in the gradient expansion of $\delta{F}$ is
given by 

\begin{equation}
  \label{eq:dos_1}
  \delta{F}_o \;=\; -4\,T\, \lim_{\varepsilon\rightarrow 0^+}
  \text{Im}\,\int_0^{\infty}\!E\,dE\, \ln\left(2\,\cosh\frac{E}{2\,T}\right)
  \left( \frac{1}{\sqrt{\Delta^2-E^2-i\,\varepsilon}} -
    \frac{1}{\sqrt{-E^2-i\,\varepsilon}} \right) + \frac{\Delta^2}{V}\;.
\end{equation}
By employing contour integration in the complex plane, it can be shown (see
Appendix~\ref{sec:ap3}) that Eq.~(\ref{eq:dos_1}) coincides precisely with
the first two terms on the RHS of Eq.~(\ref{eq:2-grad-5}).
Equation (\ref{eq:2-dos-8}) can be further simplified through integration by
parts
\begin{equation}
  \label{eq:2-dos-9}
  \int_0^{\infty}\!\frac{E\,dE}{
  \left(1-E^2-i\,\varepsilon\right)^{k+\frac{1}{2}}}\,
  \ln\left(2\,\cosh\frac{E}{2\,T}\right) \;=\; -\frac{1}{2\,T\,(2\,k-1)}
  \int_0^{\infty}\!dE\,\frac{\tanh\frac{E}{2\,T}}{
  \left(1-E^2-i\,\varepsilon\right)^{k-\frac{1}{2}}}\;.
\end{equation}
Hence
\begin{equation}
  \label{eq:2-dos-10}
  \delta{F} \;=\; \delta{F}_o +\sum_{k=1}^{\infty}
  \frac{\left\langle\delta{R}_k(x)\right\rangle}{2\,k-1}\,
  \lim_{\varepsilon\rightarrow 0^+} 
  \text{Im}\,\int_0^{\infty}\!dE\,\frac{\tanh\frac{E}{2\,T}}{
    \left(1-E^2-i\,\varepsilon\right)^{k-\frac{1}{2}}} \;.
\end{equation}
By using complex contour integration, it can be shown that (see
Appendix~\ref{sec:ap3})
\begin{eqnarray}
  \label{eq:2-dos-11}
  c_k(T) &\equiv& \lim_{\varepsilon\rightarrow 0^+}\,
  \text{Im}\,\int_0^{\infty}\!dE\,\frac{\tanh\frac{E}{2\,T}}{
    \left(1-E^2-i\,\varepsilon\right)^{k-\frac{1}{2}}} \;=\; 2\,\pi\,T
  \sum_{\omega_m>0} \left(\omega_m^2+1\right)^{-k+\frac{1}{2}}\;\nonumber\\
  & & \\
  &=& \lim_{\varepsilon\rightarrow 0^+}\,\text{Re}\,\int_0^{\infty}\!dE\,
  \frac{\tanh\frac{E}{2\,T}}{\sqrt{\tilde{E}^2-1}\,
  \left(1-\tilde{E}^2\right)^{k-1}}\;, \qquad k=2,3,\ldots \;,  \nonumber
\end{eqnarray}
where $\tilde{E}\equiv E+i\,\varepsilon$.
Finally, inserting (\ref{eq:2-dos-11}) into Eq.~(\ref{eq:2-dos-10}) leads to
our previous result (\ref{eq:2-grad-5}) and, therefore, to
Eq.~(\ref{eq:2-grad-5}) which can be also written as
\begin{equation}
  \label{eq:2-dos-12}
  \delta{F} \;=\; \delta{F}_o + \sum_{k=2}^{\infty} \frac{c_k(T)}{2\,k-1}\,
  \left\langle\delta{R}_k(x)\right\rangle \;.
\end{equation}
The coefficients $c_k(T)$ can be calculated by using their integral
representation (\ref{eq:2-dos-11}). By employing the identity $\tanh(E/2T) =
1-2\,f(E)$, where $f(E)$ is the Fermi function, one can separate $c_k(T)$
into a temperature independent and a temperature dependent part; the $T$
independent part can be calculated analytically with the result
\begin{eqnarray}
  \label{eq:2-dos-11a}
  c_k(T) &=& \lim_{\varepsilon\rightarrow 0^+}\,
  \text{Re}\,\int_0^{\infty}\!dE\,
    \frac{1-2\,f(E)}{\sqrt{\tilde{E}^2-1}\,
  \left(1-\tilde{E}^2\right)^{k-1}}
  \nonumber \\
  & & \\
  &=& \frac{2^{k-2}\,(k-2)!}{(2\,k-3)!!} \;-\; \lim_{\varepsilon\rightarrow
  0^+}\, \text{Re}\,\int_0^{\infty}\!dE\,
  \frac{2\,f(E)}{\sqrt{\tilde{E}^2-1}\,\left(1-\tilde{E}^2\right)^{k-1}}
  \;. \nonumber 
\end{eqnarray}
Furthermore, by repeated partial integration, the second term on the RHS of
Eq.~(\ref{eq:2-dos-11a}) can be expressed as an improper definite integral
involving the derivatives of the Fermi function and the familiar BCS DOS
\begin{equation}
  \label{eq:2-dos-13c}
  \rho_o(E) \;=\; \frac{E\,\Theta(E-1)}{\sqrt{E^2-1}} \;.
\end{equation}
For convenience we list below the expressions of the coefficients $c_k(T)$
for $k=2$ and $3$
\begin{equation}
  \label{eq:2-dos-13a}
  c_2(T) \;=\; 1-2\int_0^{\infty}\!\rho_o(E)\,dE\,
  \left(-\frac{\partial{f}}{\partial{E}}\right) \;=\;
  \frac{\rho_s(T)}{\rho_s(0)}\;,
\end{equation}

\begin{equation}
  \label{eq:2-dos-13b}
  c_3(T) \;=\; \frac{2}{3} -\frac{2}{3} \int_0^{\infty}\!\rho_o(E)\,dE\,
  \left(-\frac{\partial{f}}{\partial{E}}\right) -
  \frac{1}{3}\int_0^{\infty}\!\rho_o(E)\,dE\,
  \left(\frac{\partial{f}}{\partial{E}}\right)^2\;,
\end{equation}
where $\rho_s(T)$ is the superfluid density at temperature $T$.

As we have already mentioned, this second method of calculating the free
energy of an inhomogeneous superconductor by means of the effective local
density of states (\ref{eq:2-dos-6}) is quite general and in fact it is
applicable for an arbitrary distribution $f_i$ of the Bogoliubov
quasiparticles, as we show in the next section.

\subsection{Non-equilibrium Free Energy Density}
\label{sec:2-noneq}

Consider a superconducting state in which the quasiparticles are out of
equilibrium with the condensate. We also assume that the superconducting
state can be described by the effective mean--field Hamiltonian
(\ref{eq:2-BdG-1}), with a pair potential $\Delta(\bbox{r})$ and a
non-equilibrium quasiparticle distribution function $f(E;\bbox{r})$. Then,
the local DOS $\rho(E;\bbox{r})$ given by Eq.~(\ref{eq:2-dos-6a}) is
applicable with the same diagonal resolvent $R_{\pm}$ studied in the
previous sections. Thus, one can immediately write down the expressions for
the energy ($W$) and entropy ($S$) densities of the system
\begin{equation}
  \label{eq:2-noneq-1a}
  W \;=\; \frac{\left[\Delta(\bbox{r})\right]^2}{V} -
  \int_0^{\infty}\!E\,dE\, \rho(E;\bbox{r})\,\left[1 -
    2\,f(E;\bbox{r})\right] \;,
\end{equation}
and
\begin{equation}
  \label{eq:2-noneq-1b}
  S \;=\; \int_0^{\infty}\!dE\,\rho(E;\bbox{r})\,\left\{ f(E;\bbox{r})\,\ln
    f(E;\bbox{r}) + \left[1-f(E;\bbox{r})\right]\,
    \ln\left[1-f(E;\bbox{r})\right]\right\}\;.
\end{equation}
The usefulness of these equations depends on the problem at hand. For
example, if the system is in local thermal equilibrium, such that a local
temperature $T(\bbox{r})$ can be defined and the distribution function of
the quasiparticles can be expressed, e.g., as $f(E;\bbox{r}) =
\left(\exp\left[E/T(\bbox{r})\right]+1\right)^{-1}$, then it make sense to
define a free energy density through the usual thermodynamic relation
$F=W-T\,S$.
Furthermore, assuming that the considered superconducting state is close to
the equilibrium BCS state, it is straightforward to derive a gradient
expansion formula for $\delta{F}$ along the line discussed in the previous
sections.

\section{Superconductor In the Presence of the Magnetic Field and
Supercurrents}
\label{sec:finite_field}

\subsection{Free Energy}

For $\Delta ({\bf r})=|\Delta|\,e^{i\theta}$ complex, with a general spatial
dependence of the phase $\theta$, and in the presence of a static magnetic
field the squared Hamiltonian ${\cal H}_A^2 $ cannot be rotated into a
matrix with second-order differential operators on the diagonal and
off-diagonal terms equal to zero. Consequently we will go back to the
expression for the free energy (\ref{eq:2-andreev-6}), but now written as
\begin{equation}
  \label{eq:finite-1}
  {\cal F} \;=\; -2\,T\,\pi\,\hbar\,v_{\text{F}}\,N_o
  \int\!d^{\,2}r_{\perp}\! \int\!\frac{d\Omega_{\hat{\bbox{u}}}}{4\,\pi}
  \sum_{\omega_m}\ln\text{Det} \left( i\omega _m + {\cal H}_A \right) +
  \int\!d^{\,3}\bbox{r} \frac{|\Delta({\bf r})|^2}{V} + {\cal F}_H\;.
\end{equation}
Here we use the factorization $\omega_m^2+E_n^2 = \left(i\omega_m
  +E_n\right)\left(-i\omega_m+E_n\right)$, so the sum now goes over both
positive and negative Matsubara frequencies.

The determinant stays unchanged when we make the unitary transformation

\begin{eqnarray}
{\cal H}_A &\;\equiv\;&
\left(
  \begin{array}{cc}
    -i\hbar\,v_F\,\partial_x -v_F\,\frac{e}{c}\,A_{\bbox{u}} &
    |\Delta|\, e^{i\theta } \\[1.5ex]
    |\Delta|\, e^{-i\theta } &
    i\hbar\,v_F\,\partial_x -v_F\,\frac{e}{c}\,A_{\bbox{u}}
  \end{array}
\right)\;
\rightarrow \nonumber \\[2ex]
\rightarrow\;
\tilde{\cal H}_A &\equiv&
\left(
  \begin{array}{cc}
    e^{-i\theta /2} & 0 \\[1.5ex]
    0 & e^{i\theta /2}
  \end{array}
\right) \;
{\cal H}_A \;
\left(
  \begin{array}{cc}
    e^{i\theta /2} & 0 \\[1.5ex]
    0 &  e^{-i\theta /2}
  \end{array}
\right) \\[2ex]
&\;=\;&
\left(
\begin{array}{cc}
  -i\hbar\, v_F\,\partial_x + v_F\,m\,v_{s_{\bbox{u}}} & |\Delta| \\[1.5ex]
  |\Delta| & i\hbar\,v_F\,\partial_x + v_F\,m\,v_{s_{\bbox{u}}}
\end{array}
\right)\;. \nonumber
\end{eqnarray}
Here
\[
A_{\bbox{u}}\equiv {\hat{\bbox{u}}}\cdot{\bf A}\;,
\]
and
\[
v_{s_{\bbox{u}}}\;\equiv\; \hat{\bbox{u}}\cdot\bbox{v}_s \;\equiv\;
{\hat{\bbox{u}}} \cdot\frac{\hbar}{2m}\left(\nabla\theta - \frac{2e}{\hbar
    c}\,{\bf A} \right)\;,
\]
are, respectively, the components of the vector potential and the superfluid
velocity in the direction of the quasiclassical trajectory.  
In what follows, it is more convenient to use the Hamiltonian $\tilde{\cal
  H}_A $ instead of ${\cal H}_A$.

As in the absence of the magnetic field and supercurrents, the Fredholm
determinant in Eq.~(\ref{eq:finite-1}) can be calculated from the trace of
the {\em matrix\/} resolvent [cf.~Eq.~(\ref{eq:2-resolvent-7})]
\[
G(x,y;{\bf r}_{\perp }, {\hat{\bbox{u}}}; \lambda ) \;\equiv\; \left\langle
  x \left| \left(\tilde {\cal H}_A -\lambda\right)^{-1}\right| y
\right\rangle\;.
\]
We have
\[
\ln \text{Det} \left(i\omega _m + \tilde {\cal H}_A\right) \;=\; \sum_n \ln
\left(i\omega _m +E_n\right) \;=\; -\sum_n\! \int^{-i\omega _m}\!
\frac{d\lambda}{E_n-\lambda} \;=\;
\]
\begin{equation}
  \label{eq:finite-2}
  =-\!\int^{-i\omega _m} d\lambda\, \text{Tr}\, G(\lambda) \;=\; -\!
  \int^{-i\omega _m} d\lambda \int\! dx\, \text{tr}\, G(x,x;{\bf r}_{\perp
    }, {\hat{\bbox{u}}};\lambda ) \;.
\end{equation}
In Eq.~(\ref{eq:finite-2}) the notation ``$\text{Tr}$'' means the trace of
the differential operator, and so involves integration over $x$, while the
symbol ``$\text{tr}$'' means a summation over the two spin indices only. In
the following, for brevity we will omit the arguments ${\bf r}_{\perp}$ and
${\hat{\bbox{u}}}$.

As shown in Appendix~\ref{sec:ap2}, the $2\times 2$ matrix $G(x,x;\lambda )$
is obtained from the matrix function $g(x;\lambda) \equiv G(x,x;\lambda)\,
\sigma _3$, which satisfies the Eilenberger equation
\begin{equation}
  \label{Eilenberg}
  i\hbar\, v_F\, g' +
  \left[ \pmatrix{
      \lambda -v_F\,m\,{v_s}_{\bbox{u}} & -|\Delta| \cr
      |\Delta| & -\lambda + v_F\,m\,{v_s}_{\bbox{u}} \cr },
    \,g \right] \;=\; 0 \;.
\end{equation}
The Eilenberger (or quasiclassical Green's) function $g(x;\lambda)$ also
satisfies

\begin{equation}
  \text{tr}\,g \;=\; 0\;, \quad 
  g^2 \;=\; -\frac{1}{4\,\hbar^2 v_F^2}\,\sigma_0 \;.
\end{equation}
In terms of $g(x;\lambda)$, the free-energy density  is given by
\begin{equation}
  F \;=\;  2\,T\,\pi\,\hbar\,v_{\text{F}}\,N_o
  \int\!\frac{d\Omega_{\hat{\bbox{u}}}}{4\pi} \sum_{\omega_m}
    \!\int^{-i\omega_m}\! d\lambda\, \text{tr}\! 
    \left\{g(x;\lambda)\,\sigma_3\right\} + \frac{|\Delta(\bbox{r})|^2}{V} +
    F_H \;.
\end{equation}

\subsection{Gradient Expansion} 

To obtain the gradient expansion of the free energy density $F$, we rewrite
the Eilenberger equation (\ref{Eilenberg}) in the form

\begin{equation}
  \label{Eilenexp}
  i\hbar\, v_F\, g' + [V,g]+[A,g] \;=\; 0 \;,
\end{equation}
where
\[
A \;=\; \pmatrix{\lambda & -|\Delta| \cr
|\Delta| & -\lambda }.
\]
is of the zeroth order and
\[
V \;=\; \pmatrix{-v_F\,m\,v_{s_{\bbox{u}}} & 0 \cr
0 & v_F\,m\,v_{s_{\bbox{u}}}}
\]
is of the first order in gradients of the order parameter.

The contribution of the zeroth order to $g$ satisfies the equation

\begin{equation}
  \label{Eilenzero}
  [A,g_0] \;=\; 0 \;,
\end{equation}
and so it is at every point the Eilenberger function for a homogeneous
superconductor with constant real order parameter that would be equal to
$|\Delta|$ at that point. This may be calculated explicitly from the
resolvent. We find

\begin{equation}
  g_0(x;\lambda ) \;=\;
  \frac{i}{2\hbar\, v_F\, \sqrt{\lambda^2-|\Delta|^2}}\,
  \pmatrix{\lambda & -|\Delta| \cr
    |\Delta| & -\lambda} \;,
\end{equation}
which indeed satisfies Eq.~(\ref{Eilenzero}).

To obtain the higher-order terms, it is convenient to transform to the basis
of eigenvectors of $A$. That is we find a matrix $B$ such that $B^{-1} A B =
\text{diag}$. The eigenvalues of $A$ are $\pm \zeta$ with $\zeta \equiv
\sqrt{\lambda^2 - |\Delta|^2}$, and

\begin{equation}
  \label{B}
  B \;=\; \pmatrix{\lambda +\zeta & |\Delta| \cr
    |\Delta| & \lambda +\zeta } \;.
\end{equation}
A general matrix $M=M^{(0)} \sigma _0 +...+M^{(3)} \sigma _3$ (where
$\sigma_{\alpha}$, $\alpha=1,2,3$, are the Pauli matrices, and $\sigma_0$ is
the $2\times 2$ identity matrix) transforms into $\hat M \equiv \hat M^{(0)}
\sigma _0 +...+ \hat M^{(3)} \sigma _3 \equiv B^{-1}M B$, where

\begin{mathletters}
\begin{eqnarray}
  \hat M ^{(0)} & \;=\; & M^{(0)} \\
  \hat M ^{(1)} & \;=\; & M^{(1)} \\
  \pmatrix{ \hat{M}^{(2)} \cr \hat{M}^{(3)}} & \;=\; &
  \frac{1}{\zeta}\pmatrix{\lambda & i\,|\Delta| \cr
    -i\,|\Delta| & \lambda} \pmatrix{M^{(2)} \cr M^{(3)}} \;.
\end{eqnarray}
\end{mathletters}
This transformation is complex-orthogonal rather than unitary, because it
was supposed to rotate $ - i\,|\Delta|\,\sigma_2 + \lambda\,\sigma_3$ with
one component purely imaginary into $\zeta\,\sigma_3$, as it, indeed, does.

Next we define
\begin{equation}
  \label{eq:finite_R}
  R(x;\lambda ) \;\equiv\; B^{-1}\,g(x;\lambda )\,B \;.
\end{equation}
Then
\begin{equation}
  \label{eq:finite-3}
  B^{-1}\,g'\,B \;=\; R' + [B^{-1}\,B',\,R] \;.  
\end{equation}
From (\ref{B}), we obtain
\[
B^{-1}\, B' \;=\; \frac{\lambda\,|\Delta|'}{2\,\zeta^2 }\,\sigma_1
+ B_0 \;,
\]
where $B_0$ is proportional to the unit matrix $\sigma_0$ and, therefore,
contributes nothing to the commutator with $R$ in Eq.~(\ref{eq:finite-3}).
Hence, in this new basis, Eq.~(\ref{Eilenexp}) reads

\begin{equation}
  \label{dik}
  i\hbar\, v_F\, R'(x;\lambda) + [U,R(x;\lambda )] +
  \zeta\, [\sigma _3, R(x;\lambda)] \;=\; 0 \;,
\end{equation}
where $U\equiv B^{-1}\,V\,B + i\hbar\,v_F\,{(B^{-1}\,B')}^{(1)}$ is given by

\begin{mathletters}
\begin{eqnarray}
U^{(1)} & \;=\; & \frac{i\hbar\,v_F\,\lambda\,|\Delta|'}{2\,\zeta^2}\;, \\
U^{(2)} & \;=\; & -\frac{i\,|\Delta|\,v_F\,m\,v_{s_{\bbox{u}}}}{\zeta}\;, \\
U^{(3)} & \;=\; & -\frac{\lambda\,v_F\,m\,v_{s_{\bbox{u}}}}{\zeta}\;,
\end{eqnarray}
\end{mathletters}
and $R$ satisfies the conditions

\begin{equation}
 \text{tr}\, R(x;\lambda )\;\equiv\; 0\;, \quad
 R^2 \;\equiv\; -\frac{1}{4\,\hbar^2\,v_F^2}\,\sigma_0 \;.
 \label{norm}
\end{equation}
It is this matrix function $R(x;\lambda )$ that is the analogue of the {\em
  scalar} resolvent (\ref{eq:2-resolvent-1}); hence we denote it by the same
letter. The function $R$ can be expanded into the asymptotic
series\cite{dikii}

\begin{equation}
  \label{ass}
  R \;=\; \sum_{n=0}^{\infty} R_n\, \zeta^{-n} \;,
\end{equation}
where

\begin{equation}
  \label{incond}
  R_0 \;\equiv\; B^{-1}\,g_0\,B \;=\; \frac{i}{2\,\hbar\,v_F}\,\sigma_3 \;.
\end{equation}
We have somewhat generalized Dikii's work\cite{dikii}, because the expansion
parameter $\zeta^{-1}$ is $x-$dependent, and has to be, therefore,
differentiated too when we substitute (\ref{ass}) into (\ref{dik}). The
derivative of the $n-$th order term in the expansion is
\[
\left(R_n\,\zeta^{-n}\right)' \;=\; R_n'\,\zeta^{-n}
-n\,R_n\,\zeta'\,\zeta^{-n-1} \;.
\]
If we multiply $\zeta$ by a constant $C$, both terms will be multiplied by
$C^{-n}$, so they are both of the order $n$ in $\zeta^{-1}$.  Equating the
$n-$th order term in (\ref{dik}) to zero, we obtain the recurrence relation

\begin{equation}
  i\hbar\,v_F\,\left(R'_n - n\,\frac{\zeta'}{\zeta}\, R_n\right) 
  + \left[U,\,R_n\right] + \left[\sigma_3,\,R_{n+1}\right] \;=\; 0 \;.
\end{equation}
If we write $R_n = R_n^{(0)} \sigma_0+ \ldots +R_n^{(3)} \sigma_3$, then
$R_n^{(0)} = 0$ since $\text{tr}\,R = 0$, and the remaining components
satisfy the following recurrence relations

\begin{mathletters}
\label{rec}
\begin{eqnarray}
  \label{rec1}
  R_{n+1}^{(1)} & \;=\; & 
  -\frac{1}{2}\,\hbar\,v_F\,\left({R_n^{(2)}}'-n\frac{\zeta'}{\zeta}\,
  R_n^{(2)}\right) + U^{(1)}\,R_n^{(3)} - U^{(3)}\,R_n^{(1)} \;, \\
  R_{n+1}^{(2)} & \;=\; & 
  \frac{1}{2}\,\hbar\,v_F\,\left({R_n^{(1)}}'-n\frac{\zeta'}{\zeta}\,
    R_n^{(1)}\right) + U^{(2)} R_n^{(3)} - U^{(3)}\,R_n^{(2)} \;, \\
  \label{rec3}
  \left(R_n^{(3)}\,\zeta^{-n}\right)' & \;=\; &
  2\,\frac{\zeta^{-n}}{\hbar\,v_F}\, \left(U^{(2)}\,R_n^{(1)} - U^{(1)}\,
  R_n^{(2)}\right) \;. 
\end{eqnarray}
\end{mathletters}

Note that there is no recurrence relation for the coefficients $R_n^{(3)}$,
but only for the derivative of ${R_n^{(3)}\zeta^{-n}}$.  For $\zeta =
\text{const}$, the theory in Ref.~\onlinecite{dikii} guarantees that the
right-hand side of Eq.~(\ref{rec3}) is always a derivative of a polynomial
in entries of $U$ and their derivatives. The integration is, therefore,
always possible, but it leaves an undetermined constant in every
$R_n^{(3)}$. These constants together with the constants in $R_n^{(0)}$
(which are set to zero in our case since $\text{tr}\,R = 0$) determine the
solution of Eq.~(\ref{dik}) uniquely.  The product of two such solutions
again solves (\ref{dik}), so all the solutions of (\ref{dik}) form an
infinitely dimensional commutative algebra over the field of complex
numbers. Equivalently, they form a 2-dimensional algebra over the field of
formal series in $\zeta^{-1}$ with constant coefficients.

For $\zeta$ spatially dependent, a simple extension of Dikii's theory shows
that the right-hand side of (\ref{rec3}) can still be integrated; now it
will contain also powers and derivatives of $\zeta^{-1}$. However, the
spatial dependence of $\zeta$ forces all the constants on the diagonal of
$R_n$ to be zero for $n>0$, so the solution of (\ref{dik}) is completely
determined by constants on the diagonal of $R_0$. Moreover, the only
solution with $R_0=\sigma_0$ is $\sigma_0$ itself, so an arbitrary
solution can be written as
\[
R \;=\; R_0^{(0)}\,\sigma_0 + R_0^{(3)}\,\tilde{R} \;,
\]
where $R_0^{(0)}$ and $ R_0^{(3)}$ are complex numbers, and $\tilde{R}$ is
the {\em unique} solution with $\tilde{R}_0 = \sigma_3$. Hence, the spatial
dependence of $\zeta$ keeps the algebra two-dimensional, but reduces the
coefficient field from formal infinite series to complex numbers. The
algebra is therefore reduced to the two-dimensional Clifford algebra
Cl(1,C). In our case, $R=\left(i/2\hbar\,v_F\right)\,\tilde{R}$. The algebra
structure then {\em forces} $R^2= -\left(1/4\hbar^2\,v_F^2\right)
\,\sigma_0$, in agreement with (\ref{norm}).

In terms of the expansion coefficients $R^{(\alpha)}_n $ the free-energy
density has the form

\begin{equation}
  \label{fedens}
  F \;=\; 4\,T\,\pi\,\hbar\,v_{\text{F}}\,N_o
  \int\!\frac{d\Omega_{\hat{\bbox{u}}}}{4\pi}
  \sum_{\omega_m}\!\int^{-i\omega_m}\! d\lambda \sum_{n=0}^{\infty}
  \frac{i\,|\Delta|\,R^{(2)}_n(x;\lambda) + \lambda
    R^{(3)}_n(x;\lambda)}{\zeta^{n+1}} + \frac{|\Delta|^2}{V} +
  F_H \;.
\end{equation}

So, to evaluate the free energy density we just need to find the
coefficients $R_n^{\alpha}$ from the recurrence relations (\ref{rec}) with
the initial condition (\ref{incond}), substitute them into (\ref{fedens}),
and perform the Matsubara sum and the $\lambda-$ and
$\hat{\bbox{u}}-$integrations.  All the $\lambda-$integrals are of the form

\begin{equation}
  \label{ilk}
  I_{\lambda,k} \;\equiv\;
  \int^{-i\omega_m}\! d\lambda\, \frac{i\lambda}{\left(\sqrt{\lambda^2
  -|\Delta|^2}\right)^{2k+1}} \;,
\end{equation}
where $k$ is a non-negative integer.

For $k=0$, the integral diverges, and therefore needs special treatment.  If
we subtract from (\ref{fedens}) the free-energy density of a normal metal
$F_N$ then the difference of the corresponding integrals becomes finite

\begin{equation}
  \label{eq:i0}
  I_{\lambda ,0} \;=\; \int^{-i\omega_m}\! d\lambda \left( {i\lambda \over
      \sqrt{\lambda ^2 - |\Delta|^2 }} - {i\lambda \over \sqrt{\lambda ^2 }}
  \right) \;=\; \left|\omega_m\right| - \sqrt{\omega_m^2 + |\Delta|^2} \;.
\end{equation}

Using $R_0^{(3)}=i/2\,\hbar\,v_F$, we find the zeroth-order contribution to
the free energy density

\begin{equation}
  \label{F0}
  F_0({\bf r})- F_N \;=\; 4\,\pi\,T\,N_o \,\sum_{\omega_m > 0}^{\omega_D}
  \left(\omega_m - \sqrt{\omega_m^2 + |\Delta|^2}\right) +
  \frac{|\Delta|^2}{V} + F_H \;,
\end{equation}
where the spurious divergence of the infinite frequency sum can be
eliminated, as usually, by cutting it off at the Debye frequency $\omega_D$.

The first-order term $F_1(\bbox{r})$ vanishes because it contains one vector
$\hat{\bbox{u}}$ to be averaged over the unit sphere which gives zero. For
$k>1$,

\begin{equation}
  \label{eq:ik}
  I_{\lambda,k} \;=\; \frac{i^{-2k}}{2k-1}\,
  \left(\omega_m^2 + |\Delta|^2\right)^{-k+\frac{1}{2}} \;.
\end{equation}

Finally, for the average over the direction $\hat{\bbox{u}}$, we use the
symmetric integration formula

\begin{equation}
  \label{navg}
  \int\! \frac{d\Omega_{\hat{\bbox{u}}}}{4\,\pi}\,
  \left(\bbox{v}_{(1)}\cdot{\hat{\bbox{u}}}\right) \cdots
  \left(\bbox{v}_{(2k)}\cdot{\hat{\bbox{u}}}\right) \;=\; 
  \frac{\sum\limits_{\pi=\text{perm}}
    \left(\bbox{v}_{(\pi_1)}\cdot\bbox{v}_{(\pi_2)}\right) \cdots
    \left(\bbox{v}_{(\pi_{2k-1})}.\bbox{v}_{(\pi_{2k})}\right)}{(2k+1)!} \;.
\end{equation}
In this expression many of the terms will be the same. Indeed when vectors
$\bbox{v}_{(j)}$ are all different the numerator on the RHS of
Eq.~(\ref{navg}) contains only $(2k)!/k!2^{k}$ distinct terms rather than
$(2k)!$.  Note that for an odd number of vectors, the
$\hat{\bbox{u}}$-integral vanishes.

Using {\em Mathematica}, we obtained the expansion of the free energy
density functional up to the eighth order in gradients of the order
parameter. The terms are getting progressively longer, so we list them below
only up to the fourth order. To make the formula shorter, we do not perform
the $\hat{\bbox{u}}-$averaging in the fourth order, just denote it by
$\langle \ldots \rangle $ around the 21 fourth-order terms. Leaving the
explicit directional averaging to the reader has been customary in the
literature.  Also, in the fourth-order terms we write primes instead of
gradients. As an example, $\langle {\bf v}_s\,|\Delta|'\,{\bf v}_s'
\rangle$ means
\[
{1\over 15} \sum\limits_{i,j} \left({v_s}_i({\bf r})
 (\partial _i |\Delta ({\bf r})|) 
\partial _j {v_s}_j({\bf r})+
{v_s}_i({\bf r})(\partial _j |\Delta ({\bf r})|) 
\partial _i {v_s}_j({\bf r})+
{v_s}_i({\bf r})(\partial _j |\Delta ({\bf r})|)
\partial _j {v_s}_i({\bf r}) \right)
\]
In this notation, the expansion up to the fourth order reads:
\begin{mathletters}
\label{fedensexp}
\begin{equation}
  \label{fedensexp-F}
  F({\bf r}) - F_N \;=\; 4\,\pi\,T\, N_o \sum_{\omega_m > 0}^{\omega_D}
  \left(\omega_m - \sqrt{\omega _m ^2 + |\Delta|^2}\right) + {|\Delta|^2 \over
    V } + F_H + F_2({\bf r}) + F_4({\bf r}) \;,
\end{equation}
where

\begin{equation}
  \label{fedensexp-a}
  F_2({\bf r}) \;=\;   
  \frac{1}{3}\,\pi\,T\,N_o\,m^2\,v_F^2\, 
  \sum_{\omega _m} \frac{|\Delta|^2}{\left(\omega_m ^2 +
      |\Delta|^2\right)^{3/2}}\,v_s^2 +
  \frac{1}{12}\,\pi\,T\,N_o\,\hbar^2\,v_F^2\, 
  \sum_{\omega_m} \frac{\omega_m^2}{\left(\omega_m^2 +
      |\Delta|^2\right)^{5/2}}\, (\nabla|\Delta|)^2 \;,
\end{equation}
and

\begin{eqnarray}
  \label{fedensexp-b}
  F_4({\bf r}) &\;=\;& N_o\,\pi\,T\sum_{\omega _m} \left\langle
    \frac{5}{4}\,\frac{v_F^4\,m^4\,|\Delta|^4\,\bbox{v}_s^4}{\left(\omega_m^2 
        + |\Delta|^2 \right)^{7/2}} - 
    \frac{v_F^4\,m^4\,|\Delta|^2\,\bbox{v}_s^4}{\left(\omega_m^2 +
        |\Delta|^2\right)^{5/2} } -
    \frac{25}{8}\,\frac{v_F^4\,\hbar^2\,m^2\,|\Delta|^2\,\bbox{v}_s^2\,
      |\Delta|'^{\,2}}{\left(\omega_m^2+|\Delta|^2\right)^{7/2}} 
  \right.\nonumber    \\ [1ex] 
  &\;-\;&  \frac{1}{2}\,\frac{v_F^4\,\hbar^2\,m^2\,\bbox{v}_s^2\,
    |\Delta |'^{\,2}}{\left(\omega_m^2 + |\Delta|^2 \right)^{5/2}} +
  \frac{35}{8}\,\frac{v_F^4\,\hbar^2\,m^2\,|\Delta|^4\,\bbox{v}_s^2\,
    |\Delta|'^{\,2}}{\left(\omega_m^2 + |\Delta|^2 \right)^{9/2}} +
  \frac{105}{64}\,\frac{v_F^4\,\hbar^4\,|\Delta|^4\,|\Delta|'^{\,4}}{\left(
      \omega_m^2 + |\Delta|^2 \right)^{11/2} } \nonumber \\[1ex]
  &\;-\;& \frac{3}{64}\,\frac{v_F^4\,\hbar^4\,|\Delta|'^{\,4}}{\left(
      \omega_m^2 + |\Delta|^2 \right)^{7/2}} -
  \frac{49}{96}\,\frac{v_F^4\,\hbar^4\,|\Delta|^2\,|\Delta|'^{\,4}}{\left(
      \omega_m^2 + |\Delta|^2 \right)^{9/2} } - 
  \frac{5}{2}\,\frac{v_F^4\,\hbar^2\,m^2\,|\Delta|^3\,\bbox{v}_s\,
    |\Delta|'\,\bbox{v}_s'}{\left(\omega_m^2+|\Delta|^2\right)^{7/2}}
  \nonumber \\ [1ex] 
  &\;+\;& \frac{\,v_F^4\,\hbar^2\,m^2 |\Delta|\,\bbox{v}_s\,|\Delta|'\,
    \bbox{v}_s'}{\left( \omega_m^2 + |\Delta|^2 \right)^{5/2}} +
  \frac{1}{4}\,\frac{v_F^4\,\hbar^2\,m^2\,|\Delta|^2\,
    \bbox{v}_s'^2}{\left( \omega_m^2 + |\Delta|^2 \right)^{5/2}} -
  \frac{5}{4}\,\frac{v_F^4\,\hbar^2\,m^2\,|\Delta|^3\,\bbox{v}_s^2\,
    |\Delta|''}{\left( \omega_m^2 + |\Delta|^2 \right)^{7/2}}
  \nonumber \\ [1ex]  
  &\;+\;&  \frac{v_F^4\,\hbar^2\,m^2\,|\Delta|\,\bbox{v}_s^2\,
    |\Delta|''}{\left( \omega_m^2 + |\Delta|^2 \right)^{5/2}} +
  \frac{3}{16}\,\frac{v_F^4\,\hbar^4\,|\Delta|\,|\Delta|'^{\,2}\,
    |\Delta|''}{\left( \omega_m^2 + |\Delta|^2 \right)^{7/2}} -
  \frac{77}{48}\,\frac{v_F^4\,\hbar^4\,|\Delta|^3\,|\Delta|'^{\,2}\,
    |\Delta|''}{\left( \omega_m^2 + |\Delta|^2 \right)^{9/2}} 
  \nonumber \\ [1ex]
  &\;+\;& \frac{3}{16}\,\frac{v_F^4\,\hbar^4\,|\Delta|^2\,
    |\Delta|''^{\,2}}{\left( \omega_m^2 + |\Delta|^2 \right)^{7/2}} -
  \frac{1}{80}\,\frac{v_F^4\,\hbar^4\,|\Delta|''^{\,2}}{\left( \omega_m^2 +
      |\Delta|^2 \right)^{5/2}} + 
  \frac{1}{2}\,\frac{v_F^4\,\hbar^2\,m^2\,|\Delta|^2\,\bbox{v}_s\,
    \bbox{v}_s''}{\left( \omega_m^2 + |\Delta|^2 \right)^{5/2}}
  \nonumber \\ [1ex]
  &\;+\;& \left.\frac{1}{4}\,\frac{v_F^4\,\hbar^4\,|\Delta|^2\,|\Delta|'\,
    |\Delta|'''}{\left( \omega_m^2 + |\Delta|^2 \right)^{7/2}} +
  \frac{1}{40}\,\frac{v_F^4\,\hbar^4\,|\Delta|'\,|\Delta|'''}{\left(
      \omega_m^2 + |\Delta|^2 \right)^{5/2}} - 
  \frac{1}{40}\,\frac{v_F^4\,\hbar^4\,|\Delta|\,|\Delta|''''}{\left(
      \omega_m^2 + |\Delta|^2 \right)^{5/2}} \right\rangle \;.
\end{eqnarray}
\end{mathletters}
The second order term (\ref{fedensexp-a}) is identical with Werthamer's
result (Eq.~(129) in Ref.~\onlinecite{werthamer-parks}) for a clean
superconductor in finite magnetic field, and for $\bbox{v}_s=0$ this reduces
to our previous result (\ref{eq:grad-f2}).
In the same limiting case $\bbox{v}_s=0$ the expression (\ref{fedensexp-b})
of the fourth order term gives, up to a total derivative, the same result as
(\ref{eq:grad-f4a}).
The fact that the two methods we used to calculate $F_4$ are fully
independent of one another gives us confidence in the validity of our
results. However, the formula (\ref{fedensexp-b}) apparently disagrees with
the result obtained by Tewordt \cite{tewordt64-385}. 
Work is in progress to locate and understand the difference between these
two results and we hope to report our findings in this regard in a future
publication.

\subsection{Local Density of States}

As we have seen in Sec.~\ref{sec:2-dos}, an alternative route for
calculating the free energy density of an inhomogeneous superconductor is
based on the local DOS.
The free energy of a bulk superconductor can be written 

\begin{eqnarray}
  \label{fedos}
  {\cal F} & \;=\; & -2\,T\sum_{E_i\geq
    0}\ln\left(2\,\cosh\frac{E_i}{2T}\right)
  + \int\!d^3\bbox{r}\,\frac{|\Delta|^2}{V} + {\cal F}_H  \nonumber \\
  & \;=\; & -2\,\pi\,T
  \int_0^{\infty}\!dE\,\rho(E)\,\ln\left(2\,\cosh\frac{E}{2T}\right) +
    \int\!d^3\bbox{r}\,\frac{|\Delta|^2}{V} + {\cal F}_H \;,
\end{eqnarray}
where the DOS $\rho(E)$ in the quasiclassical approximation reads

\begin{equation}
  \label{eq:dos-2}
  \rho (E) = {1\over \pi}\, \text{Im} \left[ 2\, \hbar\, v_F N_o \int\!
    d^{\,2}r_{\perp }\! \int\!\! dx \int {d\Omega _{\hat{\bbox{u}}} \over
      4\,\pi} \,\text{tr}\, G(x,x; \tilde E ) \right] \;,
\end{equation}
and $\tilde E = E+i\varepsilon$.
Furthermore, Eqs.~(\ref{eq:finite_R}) and (\ref{B}) allow us to express the
trace in (\ref{eq:dos-2}) in terms of the diagonal resolvent $R(x;\lambda)$

\begin{equation}
  \text{tr} G(x,x; \tilde E ) \;=\; \text{tr}\, \left[g(x;\tilde{E})\,
    \sigma_3\right] \;=\; \text{tr}\left[B\,R(x;\tilde{E})\,B^{-1}\,
    \sigma_3\right] \;=\; 2\,\frac{i\,|\Delta|\,R^{(2)}(x;\tilde{E}) +
    E\,R^{(3)}(x;\tilde{E})}{\sqrt{\tilde{E}^2-|\Delta|^2}} \;.
\end{equation}
Hence, by employing the asymptotic expansion (\ref{ass}) the local DOS
$\rho(\bbox{r};E)$, and the corresponding free energy density $F(\bbox{r})$
can be written, respectively, in the following form

\begin{equation}
  \label{eq:dos-1}
  \rho(\bbox{r};E) \;=\; \frac{4}{\pi}\,\hbar\,v_F\,N_o\text{Im} \left[
    \int\frac{d\Omega_{\hat{\bbox{u}}}}{4\,\pi} \sum_{n=0}^{\infty}
    \frac{i\,|\Delta|\,R^{(2)}(x;\tilde{E}) +
      E\,R^{(3)}(x;\tilde{E})}{\left(
    \sqrt{\tilde{E}^2-|\Delta|^2}\right)^{n+1}} \right] \;,
\end{equation}
and
\begin{eqnarray}
  \label{eq:dos-F}
  F(\bbox{r}) &\;=\;& -8\,T\,\hbar\,v_F\,N_o\,\text{Im}\left[
  \int_0^{\infty}\!dE\,\ln\left(2\,\cosh\frac{E}{2T}\right)
  \int\frac{d\Omega_{\hat{\bbox{u}}}}{4\,\pi} \sum_{n=0}^{\infty}
    \frac{i\,|\Delta|\,R^{(2)}(x;\tilde{E}) +
      E\,R^{(3)}(x;\tilde{E})}{\left(
    \sqrt{\tilde{E}^2-|\Delta|^2}\right)^{n+1}} \right] \nonumber \\
  & & + \frac{|\Delta|^2}{V} + F_H(\bbox{r}) \;.
\end{eqnarray}

Similarly to Eq.~(\ref{fedens}), first we need to determine the relevant
coefficients $R_n^{\alpha}$ from the recurrence relations (\ref{rec}) with
the initial condition (\ref{incond}), then substitute them into
(\ref{eq:dos-F}), and finally perform the $E-$ and
$\hat{\bbox{u}}-$integrations. All the $E-$integrals are of the form

\begin{eqnarray}
  \label{iek}
  J_{E,k} & = & \text{Im} \int\limits_0^{\infty } dE \ln 2 \cosh {E\over 2T }
  {iE \over\left( \sqrt{\tilde E ^2 - \mid \Delta \mid ^2}\right)
    ^{2k+1}} = \nonumber \\
  & = & \text{Re}  \int\limits_0^{\infty } dE \ln 2 \cosh {E\over 2T }
  {E \over\left( \sqrt{\tilde E ^2 - \mid \Delta \mid ^2}\right)
    ^{2k+1}} \;.
\end{eqnarray}
In Appendix~\ref{sec:ap3} we show that the integrals defined by
Eqs.~(\ref{ilk}) and (\ref{iek}) are related through

\begin{equation}
  \label{eq:dos-ij}
  \sum_{\omega_m} I_{\lambda, k} \;=\; -{2\over \pi } J_{E,k} \;, \quad
  k=0,1,2,\ldots \;.
\end{equation}
By employing this identity, a direct comparison between Eqs.~(\ref{fedens})
and (\ref{eq:dos-F}) shows that the two routes to the free energy density
give the same result.

\section{Conclusions}
\label{sec:2-conclusion}

In this paper we have presented a general method, based on the semiclassical
limit of the Bogoliubov--de Gennes (or wave function) formulation of the
theory of weak coupling superconductivity, for calculating the (gauge
invariant) free energy density of an inhomogeneous superconductor with a
pair potential with arbitrary spatial variation and in the presence of
supercurrents and magnetic field. We have shown that the free energy density
can be expressed in terms of the diagonal resolvent of the Andreev
Hamiltonian, the semiclassical limit of the BdG Hamiltonian, which obey the
so-called Gelfand--Dikii equation. Since the solution of the Gelfand--Dikii
equation can be easily expressed in terms of an asymptotic series, our
method is most suitable for obtaining the gradient expansion of the free
energy density when the supreconducting order parameter has slow spatial
variations on a length scale set by the BCS coherence length.
To the best of our knowledge, this is the first time when the gradient
expansion of the free energy of a clean inhomogeneous superconductor, in the
general three-dimensional case and in the presence of supercurrents and
external magnetic field, has been obtained by employing the wave function
(BdG) formulation of the theory of superconductivity.
Our result for the second order term in the gradient expansion of the free
energy density coincides with the result of Werthamer
\cite{werthamer63-663,werthamer-parks} obtained more than three decades ago
by using Green's functions. 
However, our expression of the fourth order term appears to be somewhat
different from Tewordt's Green's function result \cite{tewordt64-385} and
further investigation is needed to establish the origin of this discrepancy.
Nevertheless, since in the zero-field case we have arrived at the same
result for the fourth order term in the gradient expansion of the free
energy by using two essentially different methods, we are confident in the
viability of our approach and results.

We have also shown that our method for calculating the free energy of an
inhomogeneous superconductor is applicable for states far from equilibrium
characterized by an arbitrary temperature field and quasiparticle
distribution function.

\acknowledgments

A.J.L.~would like to thank Dimple Modgil for her contribution in some early
calculations related to this work, and I.K.~thanks A.J.~Jacobs for useful
discussions.
This work was supported in part by the National Science Foundation under
grant numbers DMR91-20000 (I.K.~and A.J.L.), through the Science and
Technology Center for Superconductivity, and DMR94-24511 (\v{S}.K.~and
M.S.).
\appendix
\section{The Gelfand--Dikii Equation}
\label{sec:ap1}

Consider the one-dimensional Schr\"odinger operator
\begin{equation}
  \label{eq:ap4-1}
  \hat{H}_S(x) \;=\; -\partial_x^2 + U(x) \;, 
\end{equation}
defined on the interval $x\in[a,b]$ (any of $a$ and $b$ may be infinite), and
the associated eigenvalue problem 
\begin{equation}
  \label{eq:ap4-2}
  \hat{H}_S(x)\,\psi_n(x) \;=\; E_n\,\psi_n(x) \;,
\end{equation}
corresponding to the homogeneous boundary condition
\begin{equation}
  \label{eq:ap4-3}
  \alpha\,\psi(x) + \beta\,\psi'(x) \;=\; 0\,,\qquad \text{for}\quad x=a,b\;.
\end{equation}
Let us denote by $\psi_a$ ($\psi_b$) the solution of the equation
\begin{equation}
  \label{eq:ap4-4a}
  \hat{H}_S(x)\,\psi(x) \;=\; [-\partial_x^2+U(x)]\,\psi(x) \;=\; E\,\psi(x)
  \;, 
\end{equation}
where $E$ is an arbitrary real number, which obeys the boundary condition
(\ref{eq:ap4-3}) only at $x=a$ ($x=b$) but not at the other end of the
interval. Then the Wronskian of $\psi_a$ and $\psi_b$
\begin{equation}
  \label{eq:ap4-4}
  W(E) \;=\; \psi_a'(x)\,\psi_b(x) - \psi_a(x)\,\psi_b'(x)
\end{equation}
does not depend on $x$ and its only a function of $E$. Note that $W(E)$
vanishes only for $E=E_n$.

The Green's function $G(x,y;E)$ associated to $\hat{H}_S$ is defined through
\begin{equation}
  \label{eq:ap4-5}
  \hat{H}_S(x)\,G(x,y;E) \;=\; \hat{H}_S(y)\,G(x,y;E) \;=\; \delta(x-y)\;,
\end{equation}
and can be expressed in terms of the Wronskian (\ref{eq:ap4-4}) as
\begin{eqnarray}
  \label{eq:ap4-6}
  G(x,y;E) &=& \left\langle x\left|\frac{1}{\hat{H}_S - E}\right| y
  \right\rangle \nonumber\\
  &=& \frac{1}{W(E)}\left[\Theta(x-y)\,\psi_b(x)\,\psi_a(y) +
  \Theta(y-x)\,\psi_a(x)\,\psi_b(y)\right] \;,
\end{eqnarray}
where $\Theta(x)$ is the step function. 
One can easily check that (\ref{eq:ap4-6}) obeys Eq.~(\ref{eq:ap4-5}) and,
because by construction satisfies the boundary condition (\ref{eq:ap4-3}),
$G(x,y;E)$ is indeed the Green's function associated to $\hat{H}_S$.

The diagonal resolvent of $\hat{H}_S$ for a given energy $E$ is defined in
terms of the Green's function as
\begin{eqnarray}
  \label{eq:ap4-7}
  R(x;E) &=& \left\langle x\left|\frac{1}{\hat{H}_S - E}\right| x
  \right\rangle \nonumber\\
  &=& \lim_{\delta\rightarrow 0^+}\frac{1}{2}\left[G(x,x+\delta;E) +
  G(x+\delta,x,E)\right] \;=\;
  \frac{\psi_a(x)\,\psi_b(x)}{W(E)}\;.
\end{eqnarray}
By taking into account Eqs.~(\ref{eq:ap4-7}) and (\ref{eq:ap4-4a}), the first
two derivatives of $R_E\equiv R(x;E)$ can be written as (for brevity we drop
the arguments)
\begin{equation}
  \label{eq:ap4-8a}
  R_E' \;=\; \frac{\psi_a'\,\psi_b + \psi_a\,\psi_b'}{W}\;,
\end{equation}
and
\begin{equation}
  \label{eq:ap4-8}
  R_E'' \;=\; 2\,(U-E)\,R_E + 2\, \frac{\psi_a'\,\psi_b'}{W}\;.
\end{equation}
Then, combining Eqs.~(\ref{eq:ap4-4}), (\ref{eq:ap4-8}) and (\ref{eq:ap4-7})
we arrive at
\begin{equation}
   \label{eq:ap4-9}
   \psi_a'\,\psi_b' \;=\; \frac{W}{4\,R_E}\,\left(R_E^{\prime\,2}-1\right)\;,
\end{equation}
Finally, inserting (\ref{eq:ap4-9}) into (\ref{eq:ap4-8}) and after some
rearrangements one obtains the desired Gelfand--Dikii equation
\begin{equation}
  \label{eq:ap4-10}
  -2\,R_E\,R_E'' + R_E^{\prime\,2} + 4\,R_E^2\,(U-E) \;=\; 1\;.
\end{equation}

\section{The Matrix Gelfand--Dikii Equation}
\label{sec:ap2}

In this appendix we will study the Andreev Hamiltonian
\begin{equation}
  \hat{H}_A \;=\; -i\sigma_3\,\partial_x + \Delta\,\sigma_1
  e^{i\sigma_3\theta} \;.
\label{EQ:AndreevH}
\end{equation}
and obtain a relation, analogous to the Gelfand-Dikii equation, obeyed by
the diagonal part of its resolvent
\begin{equation}
  G_{\alpha\beta}(x,y;E) \;=\; \left\langle\alpha, x\left|
      \frac{1}{\hat{H}_A-E}\right|\beta, y\right\rangle \;.
\label{EQ:AndreevR}
\end{equation}
Here the indices $\alpha$ and $\beta$ label components in the
two-dimensional Nambu space.

To derive the Gelfand-Dikii analogue we must first write
$G_{\alpha\beta}(x,y;E)$ in a form similar to the expression we used earlier
for the resolvent $G(x,y;E)$ of the Schr{\"o}dinger Hamiltonian. Recall that
there we had
\begin{equation}
G(x,y;E) \;=\; \frac{1}{W(E)}\,\psi_a(x)\psi_b(y)
\label{EQ:SchrodingerR}
\end{equation}
where $W(E)= W\left(\psi_a,\psi_b\right)=\psi_a'\psi_b-\psi_b'\psi_a$ is
the Wronskian of the two solutions $\psi_a$, $\psi_b$ of the homogeneous
Schr{\"o}dinger equation $\hat{H} \psi_{a,b}=E\psi_{a,b}$, and is
independent of $x$.

We will begin by constructing the resolvent $G_{\alpha\beta}(x,y;E)$ for the
Andreev Hamiltonian on a finite interval $[a,b]$. If required, the limit of
an infinite domain can be taken later. To specify the problem completely we
must impose boundary conditions on the wave functions at $a$, $b$ in such a
way that the Hamiltonian is self-adjoint. This requires that the condition
for hermiticity,
\begin{equation}
  \left\langle\psi_1\left|\hat{H}_A\,\psi_2\right.\right\rangle -
  \left\langle\hat{H}_A\,\psi_1\left|\psi_2\right.\right\rangle \;=\;
  -i\left.\psi^{\dagger}_1\,\sigma_3\,\psi_2\right|_a^b \;=\; 0
\end{equation}
be satisfied in manner that treats $\psi_1$ and $\psi_2$ on an equal
footing, thus ensuring that the domain of $\hat{H}_A$ and
$\hat{H}_A^\dagger$ coincide. We impose the vanishing condition at each end
separately. Thus
\begin{equation}
  \psi^*_{1u}\psi_{2u} -\psi^*_{1l}\psi_{2l} =0, \qquad x=a,b
\end{equation}
where $u$ and $l$ refer to the upper and lower components of $\psi$
respectively. We disentangle $\psi_1$, $\psi_2$ by dividing by $\psi^*_{1u}
\psi_{2l}$ and find, for example,
\begin{equation}
  \left .\frac {\psi_{2u}}{\psi_{2l}}\right|_{x=a}= \left .\left(\frac
      {\psi_{1l}}{\psi_{1u}}\right)^*\right|_{x=a}.
\label{EQ:Beq}
\end{equation}
Eq.~(\ref{EQ:Beq}) requires that all $\psi$ in the domain of $\hat{H}_A$
obey
\begin{equation}
  \left .\frac {\psi_u}{\psi_l}\right|_{x=a} \;=\; e^{i\theta_{a}}
\end{equation}
for some real angle $\theta_{a}$. Similarly
\begin{equation}
  \left .\frac {\psi_u}{\psi_l}\right|_{x=b} \;=\; e^{i\theta_{b}} \;.
\label{Eq:BCs}
\end{equation}
These boundary angles $\theta_{a,b}$ parameterize the family of possible
self-adjoint boundary conditions. Physically one may think of them as the
phases of the order parameter of a superconductor with an infinitely large
energy-gap abutting the ends of the interval.

We now notice that, for any two solutions $\psi_1$, $\psi_2$ of the Andreev
eigenvalue problem
\begin{equation}
  \left(-i\sigma_3 \partial_x +\Delta \sigma_1 e^{i\sigma_3
      \theta}\right)\psi_{1,2} \;=\; E\psi_{1,2} \;,
\label{EQ:Andreeveq}
\end{equation}
 the quantity
\begin{equation}
  w(\psi_1,\psi_2) \equiv \psi^{\dagger}_1(x) \sigma_3\psi_2(x)
\end{equation}
is independent of $x$. To prove this, simply differentiate $w$ and use
Eq.~(\ref{EQ:Andreeveq}).  We will see that $w(\psi_1,\psi_2)$ plays the
same role for the Andreev equation as the Wronskian plays for the
Schr{\"o}dinger equation.  For example, using the boundary condition we see
that $\psi^{\dagger}_1\sigma_3\psi_1=0=\psi^{\dagger}_2\sigma_3\psi_2$.
Combining this with the constancy of $w(\psi_1,\psi_2)$, we see that $w$
must vanish identically if the two solutions $\psi_1$, $\psi_2$ are
proportional to one another. Conversely, if for two solutions of
$\hat{H}_A\psi=E\psi$ we have $w(\psi_1,\psi_2)=0$ at some point (and hence
at all points) in the interval, then
\begin{equation}
  (\psi^*_{u1}\psi_{u2} -\psi^*_{l1}\psi_{l2})|_{a} \;=\; 0
\end{equation}
or
\begin{equation}
  \left.\left(\frac{\psi_{u1}}{\psi_{l1}}\right)^*\right|_{x=a} \;=\;
  \left.\frac{\psi_{l2}}{\psi_{u2}}\right|_{x=a}.
\end{equation}
The two solutions therefore satisfy the same differential equation with the
same initial boundary condition, and so must be proportional. We have
therefore shown that $w(\psi_1,\psi_2)$ provides the same test for linear
independence as the Wronskian, $W(\psi_1,\psi_2)$.

The resolvent $G_{\alpha\beta}(x,y;E)$ will have the form
\begin{eqnarray}
  G_{\alpha\beta}(x,y;E)&\;=\;& A^L_\beta(y) \Psi^L_\alpha(x)\;,\qquad
  \text{for}\quad x<y \nonumber\\
  &\;=\;& A^R_\beta(y) \Psi^R_\alpha(x)\;,\qquad \text{for}\quad x>y
\label{EQ:GFanzatz} 
\end{eqnarray}
where $\Psi^L_\alpha(x)$, $\Psi^R_\alpha(x)$ are solutions of the
homogeneous equation and satisfy the boundary condition on the left ($L$) or
right ($R$) hand boundary respectively.  The jump-condition obtained by
integrating
\begin{equation} 
  (-i\sigma_3 \partial_x +\Delta \sigma_1 e^{i\sigma_3
    \theta}-E)\,G(x,y;E) \;=\; \sigma_o\,\delta(x-y)
\end{equation}
across the point $x=y$ is
\begin{equation} 
  i\left(\sigma_3\right)_{\alpha\alpha'} \left[\Psi^L_{\alpha'}(y)
    A^L_{\beta} - \Psi^R_{\alpha'}(y) A^R_{\beta}\right] \;=\;
  \delta_{\alpha\beta}\;.
\label{EQ:jump}
\end{equation}

To solve Eq.~(\ref{EQ:jump}) we define $W=\Psi^{\dagger\,L}\sigma_3\Psi^R$
and use the conditions $\Psi^{\dagger\,^L}\sigma_3\Psi^L=
\Psi^{\dagger\,R}\sigma_3\Psi^R=0$.  For example, on multiplying
Eq.~(\ref{EQ:jump}) by $\Psi^{\dagger\,L}_\alpha $ we find
\begin{equation}
  i\Psi^{\dagger\,L}_\alpha \left(\sigma_3\right)_{\alpha\alpha'}
  \Psi^L_{\alpha'} A^L_\beta - i\Psi^{\dagger\,L}_\alpha
  \left(\sigma_3\right)_{\alpha\alpha'} \Psi^R_{\alpha'} A^R_\beta
  \;=\; \Psi^{\dagger\,L}_\beta\;.
\end{equation}
This collapses to
\begin{equation}
-iWA^R_\beta \;=\; \Psi^{\dagger\,L}_\beta(y)\;.
\end{equation}
In this manner we obtain
\begin{eqnarray}
  G_{\alpha\beta}(x,y;E) &\;=\;& -\frac{i}{W^*}\,
  \Psi^L_\alpha(x)\Psi^{\dagger\,R}_\beta(y)\;,\qquad 
  \text{for}\quad x<y \nonumber\\
  &\;=\;& \quad\frac{i}{W}\;\Psi^R_\alpha(x) \Psi^{\dagger\,L}_\beta(y)\;, 
  \qquad \text{for} \quad x>y \;.
\label{EQ:GF}
\end{eqnarray}

Notice that $G_{\alpha\beta}(x,y;E)= G_{\beta\alpha}^*(y,x;E)$ as befits
the resolvent of a self-adjoint operator.

For the Schr{\"o}dinger problem the Gelfand-Dikii equation applies to the
diagonal $x=y$ entry in the resolvent. Our matrix-valued
$G_{\alpha\beta}(x,y;E)$ is discontinuous at $x=y$ and, as explained in
earlier sections, we must define $G_{\alpha\beta}(x,x;E)$ by taking an
average of the left and right-hand limits
\begin{equation}
  G_{\alpha\beta}(x,x;E) \;=\; \frac{i}{2}
  \left[\frac{1}{W}\,\Psi^R_\alpha(x) \Psi^{\dagger\,L}_\beta(x) -
    \frac{1}{W^*}\,\Psi^L_\alpha(x)\Psi^{\dagger\,R}_\beta(x)\right]\;.
\label{EQ:diagG}
\end{equation}
It turns out that $G_{\alpha\beta}(x,x;E)$ is not quite the most convenient
quantity to work with. Instead we use the matrix
\begin{equation}
  g_{\alpha\beta}(x;E) \;=\; G_{\alpha\beta'}(x,x;E)\,
  \left(\sigma_3\right)_{\beta'\beta}\;.
\label{EQ:Eilebergerg}
\end{equation}
The utility of this  modification is related to the coefficient of
$\partial_x$ in the Andreev equation being $\sigma_3$ instead of the
identity.

If one takes the square of the matrix $g_{\alpha\beta}$, again using
$\Psi^{\dagger\,L}\sigma_3\Psi^L= \Psi^{\dagger\,R}\sigma_3\Psi^R=0$, one
finds that
\begin{equation}
  g^2_{\alpha\beta}= \frac{1}{4}\left[-\frac{1}{W}\,\Psi^R_\alpha
    \Psi^{\dagger\,L}_{\beta'}\, \left(\sigma_3\right)_{\beta'\beta} -
    \frac{1}{W^*}\,\Psi^L_\alpha \Psi^{\dagger\,R}_{\beta'}\,
    \left(\sigma_3\right)_{\beta'\beta}\right]\;.
\end{equation}
Now
\begin{equation}
  g^2_{\alpha\beta}\Psi^L_\beta \;=\; -\frac{1}{4W^*}\,\Psi^L_\alpha
  \Psi^{\dagger\,R}_{\beta'}\, \left(\sigma_3\right)_{\beta'\beta}\,
  \Psi^L_\beta \;=\; -\frac{1}{4}\, \Psi^L_\alpha\;.
\end{equation}
Similarly
\begin{equation}
  g^2_{\alpha\beta}\Psi^R_\beta \;=\; -\frac{1}{4}\,\Psi^R_\alpha\;.
\end{equation}
Provided that $E$ is not an eigenvalue we have $W\ne0$, so the two column
vectors $\Psi^R_\alpha$ and $\Psi^L_\alpha$ are linearly independent and
together span the two-dimensional vector space at each point $x$.
Consequently these last two equations are telling us that
\begin{equation}
  g^2_{\alpha\beta} \;=\; -\frac{1}{4}\,\delta_{\alpha\beta}\;.
\label{EQ:eilenbergernorm}
\end{equation}

We also see that the trace of $g$ vanishes
\begin{equation}
  g_{\alpha\alpha} \;=\; \frac{1}{2} \left(\frac{i}{W}\,W - \frac{i}{W^*}\,
    W^*\right) \;=\; 0\;.
\end{equation}

We can therefore find three (generally complex) numbers $a_1$, $a_2$, $a_3$
such that
\begin{equation}
  a_1^2+a_2^2+a_3^2 \;=\; 1
\end{equation}
and \begin{equation} g_{\alpha\alpha} \;=\;
  \frac{i}{2}\left(a_1\sigma_1+a_2\sigma_2+a_3\sigma_3\right)\;.
\end{equation}
After inserting Eq.~(\ref{EQ:diagG}) into Eq.~(\ref{EQ:Eilebergerg}) we may
now seek the Andreev-equation analogue of the Gelfand-Dikii relation.
Because the Andreev equation is first-order, we have only to differentiate
once with respect to $x$ before we are able eliminate the $\partial_x\Psi$'s
using Eq.~(\ref{EQ:Andreeveq}). We immediately find that
\begin{equation}
  i\partial_x g \;=\; [g,\sigma_3(E -\Delta \sigma_1 e^{i\sigma_3})]\;.
\label{EQ:Eilenberegereq}
\end{equation}

The equation (\ref{EQ:Eilenberegereq}) is well-known in the
superconductivity literature as a form of the {\it Eilenberger equation}.
The present derivation is much simpler than those usually adduced. In
particular the normalization condition Eq.~(\ref{EQ:eilenbergernorm})
appears automatically and does not have to be introduced by hand.  It also
has the added advantage of demonstrating that the Eilenberger equation
should be regarded as the Andreev problem analog of the Gelfand-Dikii
equation.

Notice that the position dependent matrix $g_{\alpha\beta}(x)$ is the
diagonal part of $\left[\sigma_3\left(\hat{H}_A-E\right)\right]^{-1}$.
Equation (\ref{EQ:Eilenberegereq}) asserts that it commutes with
$\sigma_3(\hat{H}_A-E)$, i.e.,
\begin{equation}
  0 \;=\; \left[\sigma_3\left(\hat{H}_A-E\right), g\right]\;.
\end{equation}

\section{Relationship Between $I_{\lambda,k}$ and $J_{E,k}$}
\label{sec:ap3}

To prove Eq.~(\ref{eq:dos-ij}) we examine the integrals in (\ref{iek}).
Again, for $k=0$, the integral diverges. By subtracting the contribution of
the normal metal from the free-energy density, the integral takes the form:

\begin{equation}
  J_{E,0}= \text{Re} \int\limits_0^{\infty } dE \ln 2 \cosh {E\over 2T }
  \left( {E \over\ \sqrt{\tilde E ^2 - \mid \Delta \mid ^2}} -1 \right)
\end{equation}
This integral still diverges logarithmically. We can formally
integrate by parts

\begin{eqnarray}
  J_{E,0} & = & \text{Re} \left[ (\sqrt{\tilde E ^2 - \mid \Delta \mid
      ^2}-E)
    \ln 2 \cosh {E\over 2T } \right]_0^{\infty}- \nonumber \\
  & - & \text{Re} \int\limits_0^{\infty } {dE \over 2T} \tanh {E \over 2T}
  (\sqrt{\tilde E ^2 - \mid \Delta \mid ^2}-E)
\end{eqnarray}
At $E=0$, the boundary term vanishes (as $\varepsilon \rightarrow 0+$, the
contribution from the square root is pure imaginary; the second term is zero
altogether). As $E \rightarrow \infty $, the boundary term goes to $-|\Delta
| ^2 /4T$. To do the regularization consistently, we have to discard the
boundary term.  In the remaining integral, we can replace $E$ by $\tilde E$
in the argument of $\tanh $, so we can write

\begin{equation}
J_{E,0} = {1 \over 2 } \text{Re} \int\limits_{0+i\varepsilon}^{\infty +i\varepsilon}
{dE\over T} \tanh {E\over 2T }
(\sqrt{E ^2 - \mid \Delta \mid ^2}-E)
\end{equation}
We note that the integrand becomes complex conjugate upon
$E\rightarrow E^*$. Thus, by extending the lower limit of integration
to $- \infty +i\varepsilon$, we automatically obtain just the real part:
\begin{equation}
J_{E,0} \;=\; \frac{1}{4}
\int\limits_{-\infty +i\varepsilon}^{\infty +i\varepsilon}
{dE\over T}\> \tanh {E\over 2T }
(\sqrt{E ^2 - \mid \Delta \mid ^2}-E)
\end{equation}
We can now formally close the contour of integration around the imaginary
upper half axis where $\tanh(E/2T)$ has simple poles at the (positive)
fermionic Matsubara frequencies, and obtain the result:

\begin{equation}
\label{IE0}
J_{E,0} \;=\; -\pi \sum\limits_{\omega _m > 0}(\omega _m-
\sqrt{\omega _m^2 + |\Delta | ^2})
\end{equation}
This result together with (\ref{eq:i0}) yield Eq.~(\ref{eq:dos-ij}) for $k=0$.

For $k\ge 1$ everything is straightforward.  The integral $J_{E,1}$ does not
contribute because the first-order term is zero due to symmetry discussed
above. For $k>1$, $J_{E,k}$ converges. Integrating by parts and extending
the contour gives
\begin{eqnarray}
J_{E,k} & = & \text{Re} \left[ {-1 \over 2k-1 } \>
{\ln 2 \cosh {E\over 2T } \over
\left( \sqrt{\tilde E ^2 - \mid \Delta \mid ^2}\right) ^{2k-1}}
\right]_0^{\infty }+ \nonumber \\
& + & {1 \over 4(2k-1) }
\int\limits_{-\infty +i\varepsilon}^{\infty +i\varepsilon}
{dE\over T}\> {\tanh {E\over 2T } \over
\left( \sqrt{ E ^2 - \mid \Delta \mid ^2}\right) ^{2k-1}}.
\end{eqnarray}
The integrated out term vanishes. To perform the  integral we again close the
contour in the upper half plane and obtain
\begin{eqnarray}
J_{E,k} & = & {1 \over 4(2k-1) } \sum\limits_{\omega _m >0}
2\pi i {2 \over \left( \sqrt{-\omega _m^2- \mid \Delta \mid ^2}\right) ^{2k-1}}
= \nonumber \\ & = &
{\pi \over 2(2k-1)}\sum\limits_{\omega _m}
{1\over (i)^{2k-2}
\left( \sqrt{\omega _m ^2 + \mid \Delta \mid ^2}\right) ^{2k-1}}
\end{eqnarray}
By comparing this result with (\ref{eq:ik}) one can infer that
(\ref{eq:dos-ij}) holds for any positive integer $k$.



\begin{references}
\bibitem{note1}
Strictly speaking the quantity defined here is the Gibbs free energy ${\cal
  G}={\cal F}-\mu{\cal N}$, where $\mu\approx E_{\text{F}}$ is the chemical
  potential and ${\cal N}$ is the total number of particles (see the definition
  (\ref{eq:2-BdG-3}) of $H_o$). However, we continue referring to this quantity
  as ``free energy'' and calling it ${\cal F}$.
\bibitem{note2}
The situation is analogous to the eikonal approximation used in deriving the
  geometric optics limit of wave optics.
\bibitem{note3}
For example, in the vortex state of a (conventional) type-II superconductor, at
  the opposite sides of a quasiclassical trajectory which passes through the
  center of the core of an isolated vortex, sufficiently far from the vortex
  core, the pair potential has the same magnitude (equal to its bulk value) but
  opposite signs.
\end{references}
\end{document}